\documentclass[11pt]{article}
\usepackage[final]{acl}

\usepackage{times}
\usepackage{latexsym}
\usepackage[T1]{fontenc}
\usepackage[utf8]{inputenc}
\usepackage{microtype}
\usepackage{inconsolata}
\usepackage{booktabs}
\usepackage{amsmath}
\usepackage{amssymb}
\usepackage{graphicx}
\usepackage{multirow}
\usepackage{algorithm}
\usepackage{algpseudocode}
\usepackage{listings}
\usepackage{flafter}
\lstdefinestyle{stageif}{
  basicstyle=\ttfamily\scriptsize,
  columns=fullflexible,
  keepspaces=true,
  breaklines=true,
  frame=single,
  aboveskip=5pt,
  belowskip=7pt,
  captionpos=b,
  xleftmargin=0.4em,
  xrightmargin=0.4em
}

\newcommand{\needsinfo}{\texttt{NEEDS\_INFO}}
\newcommand{\ready}{\texttt{READY}}
\newcommand{\postobs}{\texttt{POST\_OBS}}
\newcommand{\completed}{\texttt{COMPLETED}}
\usepackage{xcolor}

\title{The Unreliable Progress Bar: Can LLM Agents Reliably Report\\
Task Progress Throughout Execution?}

\author{Boyang Wang\textsuperscript{1,\textdagger} \quad
Yunhan Wang\textsuperscript{2} \quad
Yalun Wu\textsuperscript{3} \\\textsuperscript{1}\,Independent Researcher \\\textsuperscript{2}\,Beihang University \\\textsuperscript{3}\,NExT++ Lab, School of Computing, National University of Singapore \\
{lolerpanda@outlook.com}
}

\begin{document}
\maketitle
\begingroup
\renewcommand{\thefootnote}{\fnsymbol{footnote}}
\footnotetext[2]{Corresponding author}
\endgroup

\begin{abstract}
Recent large language models can emit task-progress signals that agent
frameworks use to decide whether a task should continue or stop, yet
whether a model can reliably report its task progress at every stage
of a task, and where and how its reports fail, has not been studied
systematically.  We evaluate this ability on the
public benchmark $\tau^2$-bench and on StageIF, a controlled testbed in
which reporting checkpoints are placed across the task's lifecycle.
Both settings require reports at multiple task stages.  We find that
reporting reliability depends on the stage a task has reached, and that
almost every deployed model we test is reliable at some stages and
unreliable at others.  Where reporting breaks down is not the same
everywhere.  Most deployed models lose accuracy once work is under way
and recover once the task is done.  The newest generation closes that
mid-task drop and instead grows conservative at the finish line.
Our study exposes a
capability gap in task-progress reporting and provides an evaluation
protocol that spans the whole course of task execution for this ability
on which agent operation depends.  The findings indicate that agent
frameworks should not control task flow on the strength of the model's
state reports alone.
\end{abstract}
\begin{figure}[t]
\centering
\includegraphics[width=\columnwidth,keepaspectratio]{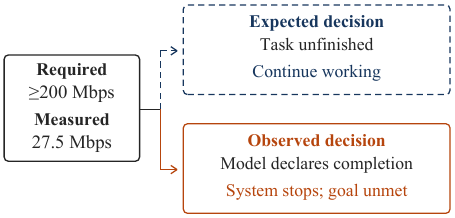}
\caption{Expected and observed decisions for the same unfinished task.
In one real \texttt{gpt} $\tau^2$-bench telecom no-user episode,
the measured speed is below the benchmark's requirement for excellent
speed.  The dashed path is the reference decision, not an executed
rollout or a claim of eventual success.  The solid path is observed:
the model calls \texttt{done()} and the runtime stops.  This native
stop signal is not an added stage tag.}
\label{fig:teaser-tau2}
\end{figure}

\section{Introduction}

Large language models increasingly run inside a loop in which a runtime
supplies context, the model acts through tools, and the cycle repeats
until someone decides the task is done
\citep{yao2023react,taubench2025,tau2bench2025,patil2025bfcl,appworld2024,wang2024mint,chen2025acebench,xu2025agentcompany}.
Runtime-owned answers to ``keep going or stop'' exist
\citep{stateflow2024,patchboard2026,fw:llamaindex,fw:semantickernel,fw:dify},
but they must be written per task.  Many agent frameworks
\citep{fw:langgraph,fw:autogen,fw:openaiagents,fw:googleadk,fw:smolagents,fw:camel,fw:metagpt,fw:claudesdk,fw:langchain,fw:autogpt,fw:vercelai,fw:mastra,fw:agno}
instead let model-generated signals participate directly in
continue-or-stop decisions (Appendix~\ref{app:survey}).  In plain terms, the model is asked to be
its own progress bar.  This paper asks whether large language models can reliably report task
progress at every stage of a task and, if not, where and how their
reports fail.  Progress here
means the task's lifecycle stage, and a report is reliable when it is
issued where the reporting duty applies and names the stage that the
environment's own state implies at that moment.  Most agent benchmarks score what an agent \emph{does}
\citep{wang2024mint,patil2025bfcl,ma2024agentboard,appworld2024,lu2025toolsandbox,wang2025steca};
this paper scores what the model \emph{tells the runtime} about whether
work remains, at moments when that value can be derived independently
from runtime state.  A structured scan of that literature found
no benchmark doing so (Appendix~\ref{app:related}).
In Figure~\ref{fig:teaser-tau2}, the measured speed is 27.5 Mbps
against a requirement of at least 200 Mbps.  The reference decision
is to continue working because the task is unfinished.  In the
observed run, however, the model declares completion and the runtime
stops while the goal remains unmet.

We first add a reporting duty to $\tau^2$-bench \citep{tau2bench2025}:
each customer-facing reply must end with a lifecycle-stage tag, scored
against the benchmark's environment state.  This separates successful
task execution from accurate progress reporting.  Some deployed
models, each a model and the serving configuration behind it, usually
omit the tag.  Among those that report, several are accurate
before acting but often name a stage the task has already passed.
Others remain accurate mid-task yet report unfinished states after
benchmark completion.  The weakness is stage-dependent, but its
location is not universal (\S\ref{sec:tau2}).

A benchmark shows where reports fail only where its dialogues happen
to place checkpoints.  StageIF, a controlled testbed, places the
reporting duty at any point of a scripted task, switches it off where
reporting would itself be an error, and freezes the correct value
before the model answers (\S\ref{sec:construct}).  It scores four
layers of failure separately: acting instead of reporting, omitting
the report, malforming it, and naming the wrong stage.  Here,
action-bearing checkpoints expose large reporting deficits, often
from omitted reports or tool calls in place of reports rather than
wrong stage values (\S\ref{sec:results}).  We call a decline at these
checkpoints the \emph{lost-mid-task pattern}; the term describes a
location, not a universal model behavior or an internal mechanism.

The two settings serve complementary purposes: natural trajectories
show the problem in task execution, while scripted checkpoints separate
report delivery, stage correctness, and correct withholding.  Bounded
interventions then test whether simple changes to the reporting demand
remove the observed deficits; they do not generally do so
(\S\ref{sec:factorial-head}).  They are checks on the findings, not a
separate claim to explain how models represent progress.  Reports do
move when task progress does not (\S\ref{sec:affordance}), and the
tested termination configurations differ in task outcome
(\S\ref{sec:realenv}).

This paper makes four contributions.
\begin{itemize}\setlength\itemsep{0pt}\setlength\parskip{0pt}
\item \textbf{Stage-dependent reporting reliability.}  We locate
  reporting weaknesses across task stages and deployed models,
  including both mid-task declines and completion-stage errors in
  models that remain accurate mid-task.
\item \textbf{Setting-dependent failure form.}  The form of failure also varies
  across evaluation settings.  Among the $\tau^2$-bench deployments
  with a mid-task decline, well-formed wrong values usually name a stage
  the task has already passed.  In StageIF, omitted reports and tool calls in place
  of reports account for much of the decline.
\item \textbf{A measurement instrument and a failure taxonomy.}
  StageIF places the reporting duty at any point in a task, freezes the
  correct value before the model answers, and scores four layers of
  failure (acting instead of reporting, omission, malformed report,
  wrong value), so both where and how a report fails become
  measurable.
\item \textbf{Bounded intervention tests.}  The tested value,
  reminder, stage-information, and reasoning changes do not generally
  eliminate the gap.  Reports change with continuation conditions
  while task progress stays fixed.  The tested termination
  configurations also yield different task outcomes.
\end{itemize}

\section{The Progress Bar on $\tau^2$-bench}
\label{sec:tau2}

\subsection{The added reporting duty}
\label{sec:tau2-setup}

$\tau^2$-bench evaluates a tool-using agent against a simulated customer
in telecom and retail domains; task success is judged by the benchmark's
own machinery---environment assertions in telecom, plus a required
transfer action on 20 hand-off tasks, and database comparison in
retail---independent of anything the agent says
\citep{tau2bench2025}.  We add one block to the agent's instructions
(Figure~\ref{fig:task-example}, Appendix~\ref{app:prompts}), requiring
every customer-facing message to end with a single tag, \needsinfo{}
while still gathering information, \ready{} once about to act with nothing
changed yet, \postobs{} once something has changed but the task is not
done, or \completed{} once nothing remains.  Tasks, tools, policies, and
the user simulator are untouched.

The correct tag at each checkpoint is derived mechanically by replaying
the conversation prefix in a fresh environment and reading the
environment's state---before the model's tag is looked at.  Environment state cannot separate
\needsinfo{} from \ready{}, so pre-action checkpoints accept either, one
resolution level below our controlled testbed; the benchmark's scripted
greeting is excluded from scoring.

\subsection{What the benchmark shows}

\begin{figure}[t]
\centering
\includegraphics[width=\columnwidth]{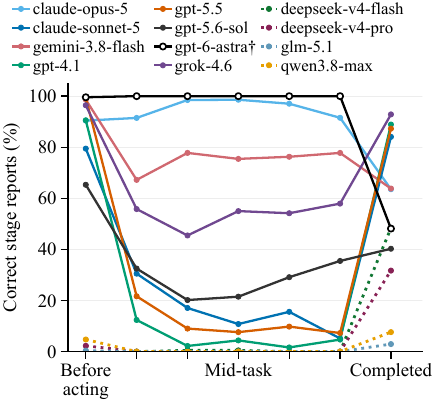}
\caption{Correct stage reports of twelve telecom deployments by the
task's true stage at the checkpoint: before acting, five equal-width
bins of mid-task position, and completed.  All curves use the strict
parser except \texttt{gpt-6-astra}$^\dagger$, whose strict counts are
$0/n$ throughout and which is drawn envelope-stripped;
\texttt{gpt-5.6-sol} stays strict.  Dotted lines mark the four
deployments that usually omit the report; the grouping describes
observed behavior, not a ranking.  \texttt{claude-opus-5} is drawn on
its 106 single-simulator trajectories.}
\label{fig:tau2-stage-all}
\end{figure}

\paragraph{First, does the deployment report?}  Of the eleven deployed
models run on the full telecom split, seven follow the duty on
93.2--100\% of checkpoints and four on only 0.6--10.1\%.
\texttt{claude-opus-5}, which ran a partial split, is the twelfth
deployment drawn in Figure~\ref{fig:tau2-stage-all}; the later figures
drop the four that usually omit the tag and read the remaining eight.
We analyze wrong-value patterns only among deployments that usually
report.

\paragraph{Where do correct reports disappear?}  \texttt{gpt-4.1}
and \texttt{gpt-5.5}, which return no reasoning tokens, are correct at
90.6--99.4\% of pre-action checkpoints and 87.3--88.9\% after completion,
but only 5.8--11.5\% mid-task.  Figure~\ref{fig:tau2-stage-all}
splits mid-task position into five bins, using trajectories with at
least two mid-task checkpoints.  For these two deployments and
\texttt{claude-sonnet-5}, accuracy falls by at least a factor of two
and a half from the first bin to the last even though the gold stage remains
\postobs{}.  These are descriptive position differences, not a
causal effect of depth.  Per-bin denominators and deployment details
are in Appendix~\ref{app:tau2-deciles}.

\paragraph{What do the wrong reports say?}
For \texttt{claude-sonnet-5}, \texttt{gpt-4.1}, and \texttt{gpt-5.5},
nearly every mid-task error carries a well-formed but wrong value, and
nine of the eleven deployments run on the full split produce no
malformed tag at all.  On the five deployments whose worst stage is
mid-task, 82--90\% of well-formed mid-task errors name a
\emph{pre-action} state already left; 10--18\% prematurely name
\completed{}.  These percentages concern wrong reports, not all
checkpoints, and do not establish delayed internal state updating.
Errors also cluster within trajectories: among those with at least
two mid-task checkpoints, 44--67\% have every mid-task report wrong
on the three collapsing deployments, versus 0--4\% entirely correct.
Such trajectories concentrate in multi-fault tasks
(Appendix~\ref{app:tau2-deciles}).

\paragraph{Parsing changes the interpretation.}  Both
\texttt{gpt-5.6-sol} and \texttt{gpt-6-astra} place stage tags inside
JSON-wrapped replies that the strict parser rejects.
Table~\ref{tab:json-envelope} separates strict scores from an
envelope-stripped diagnostic.  Stripping leaves \texttt{gpt-5.6-sol}
in a mid-task trough, whereas \texttt{gpt-6-astra} reaches 100.0\%
mid-task but only 48.1\% at completion.  Correct embedded values do
not establish strict protocol compliance; the diagnostic is not a
replacement for the main parser.

\begin{table}[t]
\centering\small
\setlength{\tabcolsep}{3pt}
\resizebox{\columnwidth}{!}{%
\begin{tabular}{@{}llrrr@{}}
\toprule
Deployment & Parser & \shortstack{Before\\acting} & Mid-task & Completed \\
\midrule
\texttt{gpt-5.6-sol} & Strict & 65.3 & 28.6 & 40.3 \\
 & Stripped & 95.5 & 42.3 & 67.5 \\
\texttt{gpt-6-astra} & Strict & 0.0 & 0.0 & 0.0 \\
 & Stripped & 99.6 & 100.0 & 48.1 \\
\bottomrule
\end{tabular}}
\caption{Correct stage reports (\%) under strict and envelope-stripped
parsing.  Stripped removes the JSON wrapper; the stage-tag parser is
unchanged.  Columns hold 245/636/77 checkpoints for
\texttt{gpt-5.6-sol} and 230/726/135 for \texttt{gpt-6-astra}.}
\label{tab:json-envelope}
\end{table}

Task success does not remove these errors: one \texttt{gpt-5.5}
trajectory succeeds despite both mid-task reports naming \ready{}
rather than \postobs{}.  Conversely, premature completion reports
occur on unfinished tasks.  Neither report accuracy nor final task
success can stand in for the other.

\paragraph{The weakness can move to completion.}
\texttt{gemini-3.8-flash}, \texttt{gpt-6-astra} read after envelope
stripping, and \texttt{claude-opus-5} on the 106 of its 114 tasks that
share one user simulator all hold 74--100\% mid-task yet only 48--64\% at
completion.  Every completion-stage wrong value of these
three names an earlier state.  Completion reporting also differs
between human-transfer tasks and other tasks, consistent with reading
the tag as ``issue resolved,'' which a hand-off is not, and with the
presence of a customer resolution cue.  Before such a cue, and outside
the hand-off tasks, the deployments that stay accurate mid-task name an
earlier state at nearly every remaining checkpoint
(Appendix~\ref{app:tau2-deciles}).  The cue
is only a keyword proxy;
these associations do not identify caution as a cause.
Appendix~\ref{app:tau2-deciles} gives the subgroup counts, reasoning
configurations, and simulator exclusions; \S\ref{sec:limits} states
their measurement limits.  None of these comparisons is a model
ranking or an isolated effect of model generation or thinking mode.

Natural dialogues locate stage-dependent weaknesses but cannot
independently place the reporting duty or distinguish the two
pre-action stages.  StageIF supplies that controlled measurement.

\section{StageIF: Measuring Lifecycle Reporting}
\label{sec:construct}

StageIF separates three questions that natural dialogues entangle:
is a report due, was it delivered, and does it name the correct stage?
It fixes checkpoint histories and gold values before generation,
so each question can be scored independently of final task success.

Consider a two-step request to move a meeting and then notify its
participants.  After the first tool succeeds, the task is incomplete and
the reporting duty is active; Figure~\ref{fig:minimal-example}
(Appendix~\ref{app:auditable-pipeline}) shows four possible model
behaviors.  Besides the correct handoff, the model
may omit the report (\emph{omission}), call the next tool instead of
reporting (\emph{divergence}), or emit a well-formed but wrong value
(\emph{wrong value}).  A malformed report is a further failure category,
distinct from a well-formed report with the wrong value.  None of these failures is
necessarily visible to an evaluation that reads only the final
environment state.  StageIF makes each of them measurable by scripting the checkpoints, so
the duty's position in the task is an experimental variable rather than an accident of dialogue
(Figure~\ref{fig:duty-position}, Appendix~\ref{app:auditable-pipeline}), and it freezes the answer key before
the model speaks.

\subsection{What is measured}
\label{sec:measured}

Three variables are kept strictly separate throughout: trusted runtime
state (what the environment records), model behavior (what the model
does and says), and trajectory outcome (whether the task ends well).
Task truth is always a function of the first; it is never inferred from
the model's report.  Deployment names are identifiers for complete
model-and-configuration bundles, not entries in a model ranking.

At each scripted checkpoint, an oracle reads runtime state alone and
answers two questions before the model's output is opened, whether a
report is due here and, if so, which of the four stage values is true.  A
deterministic parser then reads the model's reply and records whether a
well-formed report is present and what value it expresses.  Comparing
the two sides yields the paper's metrics, defined formally in
Appendix~\ref{app:formal}:

\begin{itemize}\setlength\itemsep{0pt}\setlength\parskip{0pt}
\item $\widehat\theta_z$, \emph{end-to-end adherence} (primary): at
  duty-active checkpoints of stage $z$, the share where the model spoke
  when it should, reported, and reported the true value.
\item $\widehat\phi_z$, \emph{conditional report validity} (diagnostic):
  the same, restricted to checkpoints with no assistant tool call;
  omissions and malformed reports still count as failures.  This is
  not semantic accuracy conditional on an emitted, parseable report.
\item $\widehat\omega$, \emph{correct withholding}: at duty-inactive
  checkpoints, the share where the model correctly emitted no report.
\item $\widehat\mu_t$ / $\widehat\nu_t$, \emph{false completion}: the
  share of checkpoints (respectively, of emitted reports) claiming
  \completed{} while the task is unfinished, under intervention arm $t$.
\end{itemize}

These measurements separate a reporting failure from a task failure,
since an otherwise useful reply can violate the reporting contract and a
valid report does not by itself prove task success.  Stage, dialogue
depth, and available actions change together as a task progresses, so
a stage pattern alone cannot say which one causes the gap; the
interventions below hold task truth fixed, and none establishes an
internal model representation.  Formal definitions and scoring are in
Appendix~\ref{app:formal}; evidence labels in Appendix~\ref{app:notation}.

\subsection{The instrument}
\label{sec:stageif}

We instantiate the lifecycle in 12 synthetic scenarios, divided evenly
between scheduling and customer support, each a scripted storyline that
runs from clarification through confirmation, action, reporting, and
completion.  Every scenario yields five checkpoints from a frozen history, four where
a report is due, covering all four stages, and one where the correct
behavior is a tool call with no report.  The
assistant turns inside those histories are fixed script text rather
than model output, so the checkpoints are independent of one another.

Reports are elicited in two end-positioned structured encodings, each
with a deterministic parser (Appendix~\ref{app:prompts}).  The baseline
arm requires the model to infer the stage; the gold-injected arm states
the true stage in context; the static-value arm replaces the
stage-conditioned value with a constant; a five-arm contrast later adds
inert-value and reminder conditions (\S\ref{sec:factorial}).  Seven of
nine planned deployments pass the identity checks and contribute
50{,}400 checkpoint positions, a repeated-measures count over the 12
scenarios with 20 repetitions; the exclusion criteria and audit
boundaries are in Appendix~\ref{app:admission}, and every deployment's
identifier and generation configuration is listed in
Appendix~\ref{app:models}.

\section{Where Lifecycle Reports Fail}
\label{sec:rq1}
\label{sec:results}

Does the mid-task decline persist when every stage can be identified
and the reporting duty is placed by design?  We first measure adherence
across the scripted checkpoints, then examine the failures behind the
aggregate pattern and its replication across task variants.  The next
section tests sensitivity to specific changes in the reporting
conditions, without identifying a common cause.

\subsection{Experimental Setup}
We use the baseline arm, which requires the model to infer the stage.
Scenario clusters are the statistical unit, and clustered bootstrap and
exact sign-flip tests agree on every headline decision.
Table~\ref{tab:gap} pools both encodings; $\Delta$ is the CLEAN share
among pooled nonterminal records minus the terminal CLEAN share,
with scenario-cluster bootstrap 95\% confidence
intervals over the 12 scenarios.  The all-arms pooled view, with
conditional report validity, emission, and withholding columns, is
Table~\ref{tab:gap-pooled} in Appendix~\ref{app:stats}.  Figure~\ref{fig:teaser} plots the same arm
for one structured encoding; its four checkpoints are waiting on the
user, ready to act, one step done, and task finished, ordered by task
progress rather than equally spaced in time.  Each line is one
deployment, listed alphabetically and not ranked; solid lines are the
seven originally admitted deployments, dashed lines a later cohort
never pooled with them.

\begin{figure}[t]
\centering
\includegraphics[width=\columnwidth,keepaspectratio]{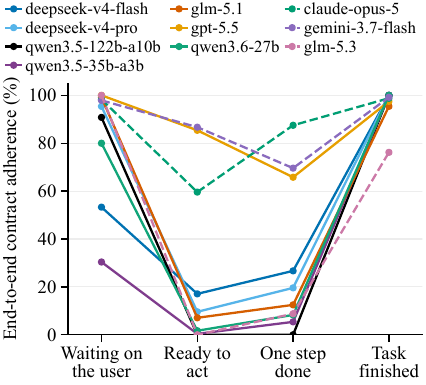}
\caption{The lost-mid-task pattern in end-to-end lifecycle-contract
adherence, one line per deployment; solid lines are the original
cohort, dashed a later one.  Accuracy drops at action-bearing
checkpoints and recovers at completion, with substantial variation
across deployments.}
\label{fig:teaser}
\end{figure}

\begin{table}[t]
\centering\small
\caption{Baseline-arm adherence $\widehat\theta$ (\%) at each
checkpoint, and the nonterminal--terminal difference $\Delta$ with its
95\% CI.  The \texttt{gpt-5.5} interval is the only one that includes
zero.}
\label{tab:gap}
\setlength{\tabcolsep}{2.2pt}
\resizebox{\columnwidth}{!}{%
\begin{tabular}{lrrrrrl}
\toprule
Deployment & Wait & Ready & Step & Done & $\Delta$ & 95\% CI \\
\midrule
deepseek-v4-flash & 31.5 & 18.8 & 15.6 & 66.2 & $-44.3$ & $[-61.7,-25.5]$ \\
deepseek-v4-pro   & 81.9 &  5.8 & 14.0 & 75.0 & $-41.1$ & $[-51.6,-29.9]$ \\
glm-5.1           & 62.5 &  4.2 &  6.7 & 59.0 & $-34.5$ & $[-43.3,-25.6]$ \\
gpt-5.5           & 57.7 & 58.2 & 34.5 & 54.4 & $-4.2$  & $[-11.5,+2.6]$ \\
qwen3.5-122b-a10b & 83.5 &  0.0 &  0.0 & 66.2 & $-38.4$ & $[-49.4,-27.9]$ \\
qwen3.5-35b-a3b   & 21.7 &  1.2 &  2.9 & 97.9 & $-89.3$ & $[-93.8,-83.9]$ \\
qwen3.6-27b       & 70.2 &  0.8 &  4.6 & 54.6 & $-29.4$ & $[-38.6,-21.7]$ \\
\bottomrule
\end{tabular}}
\end{table}

\begin{figure}[t]
\centering
\includegraphics[width=\columnwidth,keepaspectratio]{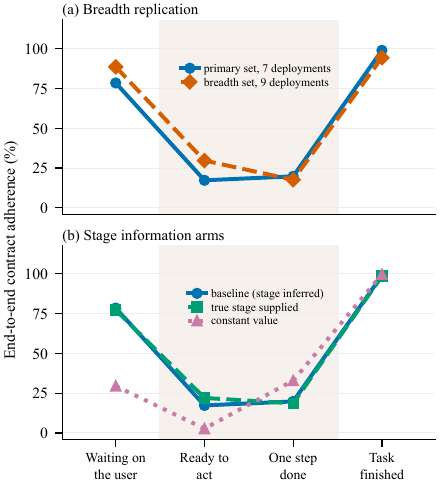}
\caption{The lost-mid-task pattern with one variable changed; blue is
the equal-weight mean of the seven original deployments.  (a) Breadth
replication, nine deployments.  (b) True-stage and constant-value
arms.}
\label{fig:valley-repeats}
\end{figure}

\subsection{Stage Pattern and Failure Modes}

\paragraph{Reporting recovers at completion.}
The single-encoding view shows lower action-bearing accuracy and
recovery at completion, not uniformly reliable reporting while
waiting.  The magnitude varies substantially.
Pooling both encodings, the nonterminal--terminal difference is
significant for six of the
seven originally admitted deployments and spans 29.4--89.3 points (Table~\ref{tab:gap}), so the same deployments
report far more reliably once the task is done.  In the all-arms pooled view, the most extreme
deployment reaches 97.4\% terminal adherence but only 8.2\% across
intermediate checkpoints (Appendix~\ref{app:stats})---nineteen in
twenty when the work is finished, fewer than one in ten while it is
happening.  The later cohort reproduces the shape with a smaller
margin, and one deployment scores lowest one checkpoint later than the
other nine, so the decline does not fall at exactly the same checkpoint
everywhere.

\begin{figure*}[t]
\centering
\includegraphics[width=\textwidth]{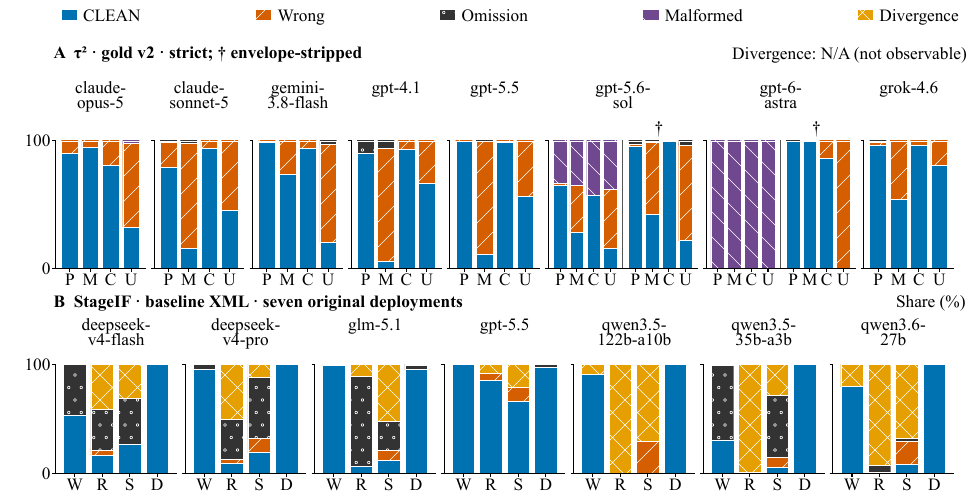}
\caption{Failure composition for eight telecom and seven original StageIF
deployments.  P/M: Pre/Mid; C/U: Completed, confirmed/unconfirmed by the
stored customer-cue keyword rule, not independently verified resolution.
W/R/S/D: Wait/Ready/Step/Done.  Telecom: strict parsing; $\dagger$ adds
envelope-stripped Sol/Astra.  StageIF: baseline XML; two transport failures
excluded.  \texttt{claude-opus-5} uses 106 single-simulator trajectories.
Bar widths do not encode sample size.  Alphabetical, not ranked;
excludes four usually-omitting telecom and three later StageIF deployments.}
\label{fig:failure-composition}
\end{figure*}

\begin{figure}[t]
\centering
\includegraphics[width=\columnwidth]{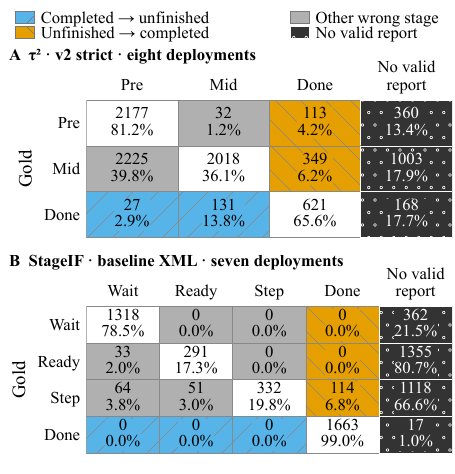}
\caption{Gold versus reported stage, pooled over eight strict-parsed
telecom and seven original baseline-XML StageIF deployments,
the same deployment set as the composition figure.  Cells give counts
and row percentages over all evaluable checkpoints, including those
with no valid report.  Telecom Pre merges \texttt{NEEDS\_INFO} and
\texttt{READY}; two StageIF transport failures are excluded.  Blue
marks an unfinished report after completion and orange a premature
completion claim.  Pooling is checkpoint-weighted and does not imply
that every deployment shows the pooled pattern.}
\label{fig:gold-matrix}
\end{figure}

\paragraph{Omission and tool calls explain much of the decline.}
Figure~\ref{fig:failure-composition} separates the outcomes at each
stage; its StageIF half is one structured encoding in the baseline
arm, across the seven original deployments at duty-active checkpoints,
read under the corrected gold on the telecom side.
Figure~\ref{fig:gold-matrix} pools gold against reported stage over
the same set.  Both count every evaluable checkpoint, so omitted and
malformed reports stay in the denominator, and both exclude two
StageIF transport failures.  Adherence falls where action
is available or pending; both missing reports and tool calls in place
of reports contribute.  Wrong values contribute too, but far less, reaching 13.6 points at the
post-action checkpoint and staying below 2 points elsewhere, so the wrong-value error that dominates on $\tau^2$-bench is
present in StageIF without dominating it.  For two deployments the
decline is almost entirely mode selection, conditional report validity
falling only a few points from terminal; for others it persists even
among turns where the model speaks (Appendix~\ref{app:stats}).  The two
settings expose different failures of the same contract, not identical
error profiles.

\paragraph{Terminal accuracy does not guarantee correct withholding.}  At
post-done checkpoints most deployments re-emit a report where emitting
is itself the error (false-alarm rates of 90--100\%;
Appendix~\ref{app:negatives}).  Reporting the right value when a final
report is due and withholding a report when none is due are separate
requirements.

\paragraph{The pattern reproduces across surface changes.}
A single-step IT-helpdesk task and an English translation reproduce
a significant gap on seven of nine deployments, and its direction on
eight and nine of nine (Appendix~\ref{app:breadth});
the two exceptions mark heterogeneity, not a ranking.
Figure~\ref{fig:valley-repeats}(a) uses the same axes and encoding as
Figure~\ref{fig:teaser} on a separate scenario set, sharing seven
deployments with the primary set and adding two.  It replicates
scenarios and surface forms, not an independent deployment
population.  Because the farthest checkpoint
recovers, simple decay with distance from the instruction does not
predict the shape; but stage, position, and action context still
co-vary, so we next change the reporting conditions while holding the
scripted task fixed.

\section{Bounded Tests of Reporting Interventions}
\label{sec:factorial-head}
\label{sec:factorial}

The core finding concerns where and how reports fail.  These checks
ask whether specific changes to the reporting demand remove the
deficit, not whether a single mechanism explains it.

\paragraph{Changing the required report.}
Supplying the true stage or replacing it with a constant does not
generally close the gap (Figure~\ref{fig:valley-repeats}(b);
Appendix~\ref{app:stats}, Figure~\ref{fig:interventions}).  For the
gold-injected arm the upper confidence limits cover at most one sixth
of the observed gap on the six deployments whose gap is significant.
In a separate five-arm batch, a trailing reminder reduces omission
but leaves tool-call divergence largely unchanged; its largest gain is
$+9.3$ points pooled over all five checkpoints, against gaps of
29--89 points.  A matched
reasoning-mode contrast redistributes failures rather than removing
them (Appendix~\ref{app:thinking}).  These tests concern particular
prompts and configurations; they do not rule out memory or reasoning
as contributing factors.

\paragraph{Comparing duties on the same histories.}
\label{sec:bridge}
Three matched obligations, with and without reminders, are evaluated
on nine deployments (Appendix~\ref{app:bridge},
Table~\ref{tab:bridge}).  A final-reply duty reaches 99.0\%, compared
with 43.7\% and 57.4\% over the duty-active checkpoints of a status
sentence and a machine report.  All three are near ceiling at the
terminal checkpoint; the machine report reaches only 43.7\% at
intermediate checkpoints.  Thus the contrast is not simply an
inability to emit a structured answer.  Fluent replies can omit the
report, unlike content truncation under a valid schema discussed by
\citet{fan2026capacity} (Appendix~\ref{app:metrics}).

\paragraph{Changing the continuation context.}
\label{sec:affordance-head}
\label{sec:affordance}
Jointly withdrawing a needed tool and assigning the unfinished step
to another system raises false completion from 6.2\% to 64.4\%
across nine deployments, despite unchanged task truth.
This is a bundled intervention, not the effect of tool withdrawal
alone.  Its factorial follow-up and per-deployment estimates remain
in Appendix~\ref{app:affordance}; neither establishes an internal
explanation for the stage pattern.

\section{Operational Consequences}
\label{sec:realenv}

Reporting and termination design also change task outcomes.  In a
validating rollout where task success is read from environment state
(12 scenarios, six repetitions, eight deployments;
Appendix~\ref{app:gate}), the marker-reporting configuration completes
13.0 points fewer tasks than the no-tool-call configuration
(scenario-clustered CI $[+7.1,+19.4]$).  The two arms match user
messages but not the full model input, because only the marker arm
carries the reporting instructions, so this contrasts
reporting-and-termination configurations rather than isolating a
gate rule.  Premature completion stops occur on 1.6\% of marker
trajectories and missing-marker stops on 22.7\%; these frequencies
describe how sessions ended and do not decompose the task-success
difference causally.  This rollout is not evidence that the
$\tau^2$-bench mid-task wrong values caused task failures.  An
independent audit also examines termination reliability
\citep{advani2026}.

\section{Related Work}
\label{sec:related}

\paragraph{Instruction following and structured output.}
IFEval \citep{zhou2023ifeval} and its multi-turn successors
\citep{he2024multiif,laban2025lost,sirdeshmukh2025multichallenge,han2025entangled,li2025structflow,wang2026evolif},
AgentIF's conditional constraints in long agent prompts
\citep{agentif2025}, and structured-output studies that separate
content from realization
\citep{geng2025jsonschemabench,gu2025structeval,lee2026formattax,shen2025slot,yuan2026structuredcausal}
evaluate user-facing response constraints; our target is a
stage-dependent value consumed by the runtime
(Appendix~\ref{app:related}, Table~\ref{tab:construct-map}).

\paragraph{Progress estimation and state-based evaluation.}
Multi-turn tool benchmarks
\citep{wang2024mint,patil2025bfcl,chen2025acebench,liu2025dialogtool},
progress-exposing analysis boards \citep{ma2024agentboard,lumina2026},
programmatic-state environments
\citep{appworld2024,lu2025toolsandbox,taubench2025}, and process
evaluators and scaffold audits
\citep{wang2025steca,zhang2026planreward,proxystate2026,sopbench2025,octobench2026}
score actions and milestones, not the model's own report of where the
task stands; $\tau^2$-bench's no-user mode consumes a model-emitted
\texttt{done} without scoring it \citep{tau2bench2025}, and
Appendix~\ref{app:tau2-deciles} records zero-reward endings on up to 63 of 114
telecom tasks per model.  Closer work does score reports.  Per-step progress and completion
estimates of UI agents against human
annotation, fed back to the agent's own planner
\citep{bishop2024latent}; RePro's progress percentages, which lack
per-step truth in outcome-based tasks and hurt performance when
prompted online \citep{repro2026}; a self-verdict loop that accepts
stagnation as improvement \citep{park2026mirage}; and terminal
reports, action claims, and completion claims against hidden world
state, execution traces, or the assigned goal
\citep{vigil2026,cao2026corrupt,arike2025goaldrift,advani2026,deploybench2026,handbook2026}.
None anchors the report to environment-derived state under a duty that
is active at some checkpoints and forbidden at others, so none
separates a wrong stage value from an omitted report or compares
reliability across stages.  The shape resembles position-sensitive
degradation \citep{liu2024lostmiddle}, but nothing moves within the
context, the terminal checkpoint is farthest from the instruction yet
recovers, and the collapse tracks whether an action is available; and
unlike unfaithful chain-of-thought explanations \citep{turpin2023},
the required value is derived from runtime state before the response
is read.  We validate oracle and verifier with negative controls and
corruption tests \citep{rigorousbench2025}.

\paragraph{Prospective memory and runtime-owned control.}
Prospective-memory studies test whether a model executes a delayed
obligation at its cue and find that reminders repair some omissions
\citep{mittal2026forget,pmbench2026,triggerbench2026}; the matched
terminal obligation in \S\ref{sec:bridge} tests the overlap directly.  Explicit workflow state \citep{stateflow2024,patchboard2026},
protocol comparisons \citep{protocolbench2025}, verifier-paired early
exit \citep{lu2025runaway}, open-source runtimes that mix
model-emitted, tool-derived, and runtime-owned termination rules
\citep{fw:langgraph,fw:autogen,fw:openaiagents,fw:googleadk,fw:llamaindex,fw:semantickernel,fw:pydanticai,fw:openhandssdk},
and a static analysis of 6{,}549 agent repositories that does not
count a model-dependent exit as a bound \citep{hou2026ial}
(Appendix~\ref{app:survey}, Table~\ref{tab:framework-signal-map})
motivate a design alternative but do not measure the reliability of a
delegated lifecycle signal.

\section{Conclusion}

Can models reliably report progress throughout a task?  In the tested
settings, neither final task success nor an accurate terminal report
establishes reliable reporting along the way, and several deployed
models name earlier stages while work is still under way.  The newest
generation of one family closes the mid-task collapse and grows
conservative at the finish line.  StageIF separates an absent report,
or one replaced by a tool call, from one whose value is wrong, and
simple changes to the reporting demand did not remove the deficit.
Measuring it therefore means scoring every stage and keeping delivery
apart from correctness.  Where independent task state exists, a
progress report should be checked against it, not treated as sole
control authority.

\section*{Limitations}
\label{sec:limits}

Each deployment
identifier names a model-and-serving bundle, newer endpoints may
silently drop sampling parameters, and no value is a ranking.  The
bundled intervention identifies a total effect, all effects are
behavioral and identify no internal mechanism, and rollout terminal
checkpoints are selected by survival, so they identify no rollout stage
gap.

\section*{Ethics Statement}
The motivating seed artifacts are
stored in desensitized form and no user identifiers are intentionally
included; any public release remains subject to a separate privacy review.
Measurements
characterize deployments at a point in time and are unsuitable for
vendor comparison or procurement decisions; we deliberately present no
ranking.  The 25 current run summaries that expose per-deployment token usage record 111,998,943 prompt-plus-completion tokens for successful responses; runs without that field and the usage of failed attempts are excluded, so this is a recomputable lower bound rather than total project cost.  Artifact release is planned only
subject to owner, privacy, licensing, and venue review.

\bibliography{custom}

\appendix
\section{Formal definitions and estimators}
\label{app:formal}

Task truth is a
function of trusted runtime state; model behavior and model-reported
state are separate variables; and trajectory outcome is never inferred
from the model's report.  Deployment names are identifiers for complete
model-and-configuration bundles, not entries in a model ranking.

Let a typed trajectory and checkpoint be
\begin{equation}
\label{eq:trajectory-checkpoint}
 \tau_i=(e_{i,1},\ldots,e_{i,T_i}), \qquad c=(i,t).
\end{equation}
Equation~\ref{eq:trajectory-checkpoint} indexes trusted runtime state
$X_c$, model-visible history $H_c$, and trajectory outcome $K_i$.
$Y_c=(B_c,R_c)$ is the model output.  The interaction-mode variable
$B_c$ records whether the model speaks or invokes a tool; $R_c$ is raw
assistant text.  Table~\ref{tab:notation} (Appendix~\ref{app:notation}) summarizes all symbols.  A frozen oracle maps
runtime state to
\begin{equation}
\label{eq:oracle}
  G(X_c)=(D_c,Z_c^*),
\end{equation}
where $D_c\in\{0,1\}$ says whether the reporting duty is active and,
when $D_c=1$, $Z_c^*\in\mathcal Z$ is the lifecycle truth, with
$\mathcal Z=\{\needsinfo,\ready,\postobs,\completed\}$; the expected
reported value is $L_c^*=g(Z_c^*)$.  Algorithm~\ref{alg:score} scores
one checkpoint.  The oracle in Equation~\ref{eq:oracle} uses only
script-declared tool status, slot
completeness, confirmation state, and remaining steps, and is evaluated
before the model output $Y_c$ is read (line~1 versus line~3); a
deterministic parser $\rho(R_c)$ then reports presence,
well-formedness, uniqueness, final position, and the expressed value
$\widehat L_c$.  Unlike conditional action constraints
\citep{octobench2026,agentif2025},
this obligation produces a value consumed by the runtime, and the runtime
can independently derive both its applicability and truth.  Let $S_c=1$
indicate that the test obtained a valid model response; all empirical
model-behavior denominators below are subsets of $\{c:S_c=1\}$, and
transport failures are reported separately and never scored as model
behavior.

For a duty-active checkpoint, $M_c$ and $E_c$ are tested at lines 6 and
8; parser-derived realization $F_c$ is tested at line 9; and value
equality $V_c=[\widehat L_c=L_c^*]$ is defined and tested at lines 10--11
of Algorithm~\ref{alg:score}.  These are respectively the required
interaction mode, report presence, valid realization, and value equality
with $L_c^*$; $E_c=0$ implies
$V_c=F_c=0$.  We report one primary metric and one diagnostic rather
than hiding two failure modes in one number:

The artifact repository (Appendix~\ref{app:repro}) grounds this mapping with three
searchable listings: a frozen case excerpt, the independently derived lifecycle truth, and a stored raw response with mechanically recomputed
parser/classifier claims.  They show concretely that the answer, answer
key, and scoring rule are separate objects.
Figure~\ref{fig:auditable-pipeline} (Appendix~\ref{app:auditable-pipeline}) places this checkpoint-level comparison inside the complete evidence path, from a hash-frozen scenario contract to recomputable aggregate metrics.

\begin{equation}
\label{eq:theta}
 \theta_z=\Pr(M_cE_cV_cF_c=1\mid D_c=1,Z_c^*=z)
\end{equation}
is \emph{stage-conditioned control adherence}, the primary end-to-end
runtime-facing metric, while
\begin{equation}
\label{eq:phi}
 \phi_z=\Pr(E_cV_cF_c=1\mid D_c=1,Z_c^*=z,M_c=1)\!
\end{equation}
is \emph{conditional report validity}, a secondary diagnostic conditional on
the checkpoint carrying no assistant tool call.  Thus
Equations~\ref{eq:theta} and~\ref{eq:phi} differ only in whether
interaction-mode selection is part of the outcome or the conditioning
set.  Action divergence lowers $\theta_z$ but lies outside $\phi_z$; report
omission, wrong value, and malformed realization lower both.  For
\begin{equation}
\label{eq:successful-stage-set}
 \mathcal C_z=\{c:S_c=1,D_c=1,Z_c^*=z\},
\end{equation}
their empirical counterparts are
\begin{align}
\label{eq:theta-hat}
 \widehat\theta_z={}&\frac{\sum_{c\in\mathcal C_z}M_cE_cV_cF_c}{|\mathcal C_z|},\\
\label{eq:phi-hat}
 \widehat\phi_z={}&\frac{\sum_{c\in\mathcal C_z}M_cE_cV_cF_c}
         {\sum_{c\in\mathcal C_z}M_c}.
\end{align}
Equations~\ref{eq:theta-hat} and~\ref{eq:phi-hat} instantiate the two
population quantities over the successful-response set in
Equation~\ref{eq:successful-stage-set}; $\widehat\phi_z$ is undefined if
its denominator is zero.  At $D_c=0$, with
$\mathcal C_0=\{c:S_c=1,D_c=0\}$, correct withholding is
\begin{equation}
\label{eq:withholding}
 \widehat\omega=|\mathcal C_0|^{-1}
 \sum_{c\in\mathcal C_0}(1-E_c).
\end{equation}
False alarms are the complement of Equation~\ref{eq:withholding}.
Task outcome $K_i$ is a parallel track, never a substitute: a useful clarification with a missing lifecycle report can be a task success and protocol
failure simultaneously.

For interventions that may induce a false completion report, define
\begin{equation}
\label{eq:false-completion-indicator}
 W_c=\mathbf 1[\widehat L_c=\completed\land L_c^*\ne\completed],
\end{equation}
with $W_c=0$ when no report is emitted.  For treatment arm $t$ over its
successful-response set $\mathcal C_t$, we distinguish
\begin{align}
\label{eq:mu-hat}
 \widehat\mu_t={}&\frac{\sum_{c\in\mathcal C_t}W_c}{|\mathcal C_t|},\\
\label{eq:nu-hat}
 \widehat\nu_t={}&\frac{\sum_{c\in\mathcal C_t}W_c}
                         {\sum_{c\in\mathcal C_t}E_c}.
\end{align}
Equation~\ref{eq:mu-hat} has a fixed successful-checkpoint denominator;
Equation~\ref{eq:nu-hat} conditions on the post-treatment event that a report was
emitted and is therefore descriptive, not a separate causal
effect.  Equation~\ref{eq:false-completion-indicator} supplies the common
numerator event.

Each result carries one of four evidence labels (\textsc{Observational},
\textsc{Controlled intervention}, \textsc{Post-treatment descriptive},
\textsc{Audit/engineering}), defined in Appendix~\ref{app:notation};
none by itself establishes an internal model representation or
intention.  A fixed priority assigns one error code per checkpoint, the
scientific split being between silent report omission and action
divergence, where the model calls a tool instead of speaking; both
violate the runtime-facing contract, although divergence also borders
agent-policy adherence \citep{sopbench2025} (Appendix~\ref{app:taxonomy}).
Finally, the oracle fixes reachable $(D_c,Z_c^*)$ pairs: only
\needsinfo{} is derivable both before and after consequential action,
\ready{} precedes it, and \postobs{} and \completed{} follow it, so
stage and dialogue depth are construct-confounded rather than merely
sample-confounded, and the stage pattern is descriptive unless a
separate intervention holds this structure fixed.  An exploratory replay
of the production protocol motivated the instrument but could not
separate stage, prompt, vocabulary, domain, or serving explanations
(Appendix~\ref{app:seed}).

\paragraph{Intervention and gate estimands.}  For the
withdraw-and-delegate bundle of \S\ref{sec:affordance}, let
$T_c^B=1$ denote the complete bundle and $T_c^B=0$ its matched
available/self-responsible arm, and let $E_c(t)$ and $W_c(t)$ denote
report emission and unconditional false completion under assignment
$t\in\{0,1\}$.  The target quantities are the complete-bundle contrasts
\begin{align}
\label{eq:bundle-emission-effect}
 \Delta_E^B={}&\mathbb E[E_c(1)-E_c(0)],\\
\label{eq:bundle-false-completion-effect}
 \Delta_\mu^B={}&\mathbb E[W_c(1)-W_c(0)].
\end{align}
The identified contrast is the bundle's total effect, not either
component's effect.  For the fully crossed follow-up, the factor assignment and deployment-specific cell means are
\begin{align}
\label{eq:factor-assignment}
 \mathbf T_c &=(T_c^O,T_c^X,T_c^C)\in\{0,1\}^3,\\
\label{eq:factor-cell-mean}
 m_{d,J}(\mathbf t)&=\mathbb E[J_c(\mathbf t)\mid d].
\end{align}
Equations~\ref{eq:factor-assignment} and~\ref{eq:factor-cell-mean}
define the crossed cells and their means, where
$J\in\{M_cE_cV_cF_c,E_c,W_c\}$.  Averaging the cell means in
Equation~\ref{eq:factor-cell-mean}, the reported marginal factor contrast is
\begin{equation}
\label{eq:factor-contrast}
 \delta^{(j)}_{d,J}=\frac{1}{4}\sum_{\mathbf t_{-j}}
 \left[m_{d,J}(1,\mathbf t_{-j})-m_{d,J}(0,\mathbf t_{-j})\right].
\end{equation}
Equation~\ref{eq:factor-contrast} averages each main effect over the
other two crossed factors.
For the validating rollout of Appendix~\ref{app:gate}, let $g$ denote
the full reporting-and-termination configuration, rather than the gate
rule alone.  With potential trajectory outcome $K_i(g)$, the target
configuration contrast is
\begin{equation}
\label{eq:gate-task-effect}
 \Delta_K^{B2-B1}=\mathbb E[K_i(B2)-K_i(B1)].
\end{equation}

\section{Error taxonomy}
\label{app:taxonomy}

Each checkpoint receives exactly one code by priority:
\texttt{TRANSPORT} (no complete response; never counted against the
model) $>$ \texttt{APP\_MISS} (spoke without protocol) /
\texttt{APP\_DIVERGENCE} (issued a tool call where the stage demanded
speaking) / \texttt{APP\_FALSE\_ALARM} (protocol on a $D_c{=}0$ turn)
$>$ \texttt{MALFORMED} $>$ \texttt{STATE\_WRONG} $>$ \texttt{CLEAN}.
Thus a parseable wrong value is assigned \texttt{STATE\_WRONG} only when
the report is otherwise well formed, unique, and final; malformed syntax or
placement is assigned \texttt{MALFORMED} even if a candidate value can be
recovered.
The stored evaluator's fine-grained \texttt{STATE\_WRONG\_VALUE} code
is normalized to \texttt{STATE\_WRONG} in this paper.
\texttt{APP\_MISS} and \texttt{APP\_DIVERGENCE} are deliberately
separated, reflecting a two-level structure: stage recognition feeds
two distinct downstream obligations---a \emph{speech/action policy}
(should this turn speak or act?), whose violation is
\texttt{APP\_DIVERGENCE}, and a \emph{protocol obligation} (when
speaking, emit the marker), whose violation is \texttt{APP\_MISS} and
is a report-validity failure.

\section{Notation and evidence map}
\label{app:notation}

\paragraph{Evidence labels.}  We use
\textsc{Observational} for conditional distributions without an
intervention; \textsc{Controlled intervention} only for the total effect
of the treatment actually changed; \textsc{Post-treatment descriptive}
for quantities such as $\widehat\nu_t$ that condition on a treatment-induced
event; and \textsc{Audit/engineering} for claims established from runtime
order, source, or traces.  These labels bound interpretation: none by
itself establishes an internal model representation or intention.

Table~\ref{tab:notation} is the semantic source of truth for the paper's
symbols; Table~\ref{tab:evidence-map} maps each study to its primary
question and interpretation boundary.

\begin{table*}[t]
\centering\small
\caption{Semantic source of truth for the paper.  Prose and figures are
constrained projections of these objects.}
\label{tab:notation}
\setlength{\tabcolsep}{5pt}
\begin{tabular}{p{0.16\textwidth}p{0.25\textwidth}p{0.51\textwidth}}
\toprule
Symbol & Source/type & Meaning \\
\midrule
$\tau_i,c=(i,t)$ & typed trace & trajectory and evaluable checkpoint \\
$X_c,H_c$ & runtime / model-visible & trusted runtime state and the history visible to the model \\
$Y_c=(B_c,R_c)$ & model response & interaction mode and raw assistant text \\
$G(X_c)=(D_c,Z_c^*)$ & frozen oracle & whether reporting is required and the lifecycle truth \\
$L_c^*,\widehat L_c$ & oracle / parser & required report value and parsed model-reported value \\
$S_c,M_c,E_c,V_c,F_c$ & binary indicators & valid model response, required mode, report presence, value correctness, and valid realization \\
$\theta_z,\phi_z$ & population estimands & primary end-to-end adherence and conditional diagnostic validity \\
$\widehat\theta_z,\widehat\phi_z$ & empirical estimators & corresponding estimates on successful model-response checkpoints \\
$K_i$ & environment/evaluator & trajectory-level task outcome, separate from protocol adherence \\
$U_c,Q_c$ & model-visible action context & task-advancing tool set and $\mathbf 1[|U_c|>0]$ \\
\bottomrule
\end{tabular}
\end{table*}

\begin{table*}[t]
\centering\scriptsize
\caption{Evidence map.  Each study has one primary question and an
explicit interpretation boundary.  Detailed arms, denominators, and
robustness analyses are in the appendix.}
\label{tab:evidence-map}
\setlength{\tabcolsep}{3pt}
\begin{tabular}{p{0.08\textwidth}p{0.18\textwidth}p{0.19\textwidth}p{0.15\textwidth}p{0.24\textwidth}}
\toprule
Step & Design & Primary outcome & Evidence type & Interpretation boundary \\
\midrule
Stage & Frozen stage checkpoints and breadth replications & $\widehat\theta_z$, $\widehat\phi_z$, failure mode & Descriptive & Stage and trajectory position co-vary \\
Alternatives & Gold/static/reminder/thinking contrasts and matched obligations & Adherence under tested alternatives & Controlled contrasts & Bounds tested accounts; not an exhaustive mechanism test \\
Context & Matched bundle plus fully crossed ownership, executability, and cue follow-up & End-to-end adherence; emission and false completion diagnostics & Interventions & Original bundle is joint; follow-up identifies explicit interface fields, not internal state \\
System & Validating rollout with alternative termination designs & Task success and termination outcome & Configuration contrasts & Matching user turns does not establish identical system prompts; replay details in Appendix~\ref{app:gateab} \\
\bottomrule
\end{tabular}
\end{table*}

\section{From Frozen Scenario to Auditable Score}
\label{app:auditable-pipeline}

\begin{figure}[h]
\centering
\includegraphics[width=\columnwidth]{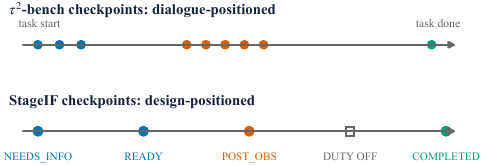}
\caption{Where the reporting duty is observed.  In $\tau^2$, checkpoints
cluster at a few stages and few occur mid-task; StageIF places or disables
the duty by design.  The duty is disabled on tool-call turns.}
\label{fig:duty-position}
\end{figure}

\begin{figure}[t]
\centering
\includegraphics[width=\columnwidth,height=0.55\textheight,keepaspectratio]{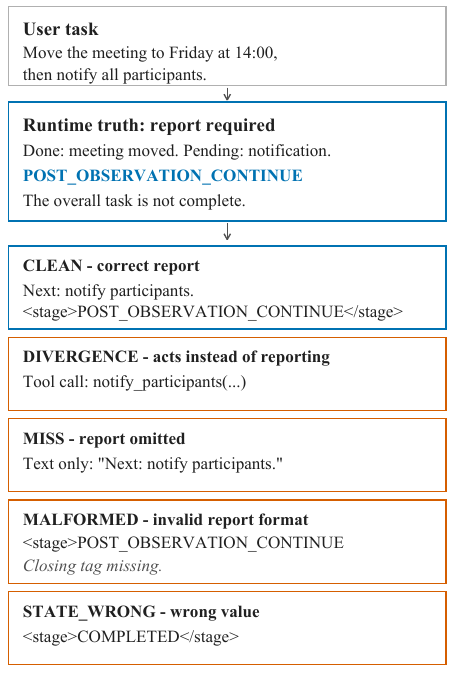}
\caption{Five model behaviors at the same duty-active checkpoint.  Only
the correct handoff supplies a machine-readable lifecycle value that
matches runtime-derived truth.  Only the divergence example includes a
tool call.}
\label{fig:minimal-example}
\end{figure}

\begin{algorithm}[t]
\caption{Scoring one checkpoint $c$.  Truth is fixed before the model
output is read; exactly one code is assigned by the frozen error-code
priority.}
\label{alg:score}
\begin{algorithmic}[1]
\State $(D_c,Z_c^*)\gets G(X_c)$;\; $L_c^*\gets g(Z_c^*)$ \Comment{runtime state only}
\State \textbf{if} no valid response \textbf{then return} \texttt{TRANSPORT} \Comment{$S_c{=}0$; no denominator}
\State $(B_c,R_c)\gets Y_c$ \Comment{model output read only now}
\State $(E_c,F_c,\widehat L_c)\gets\rho(R_c)$ \Comment{deterministic parser}
\State \textbf{if} $D_c{=}0$ \textbf{then return} \texttt{APP\_FALSE\_ALARM} if $E_c{=}1$ else \texttt{CLEAN}
\State $M_c\gets[\,B_c=\textsf{speak}\,]$
\State \textbf{if} $M_c{=}0$ \textbf{then return} \texttt{APP\_DIVERGENCE}
\State \textbf{if} $E_c{=}0$ \textbf{then return} \texttt{APP\_MISS}
\State \textbf{if} $F_c{=}0$ \textbf{then return} \texttt{MALFORMED}
\State $V_c\gets[\,\widehat L_c=L_c^*\,]$
\State \textbf{if} $V_c{=}0$ \textbf{then return} \texttt{STATE\_WRONG}
\State \textbf{return} \texttt{CLEAN} \Comment{$M_cE_cV_cF_c=1$}
\end{algorithmic}
\end{algorithm}

\begin{figure*}[t]
\centering
\includegraphics[width=\textwidth]{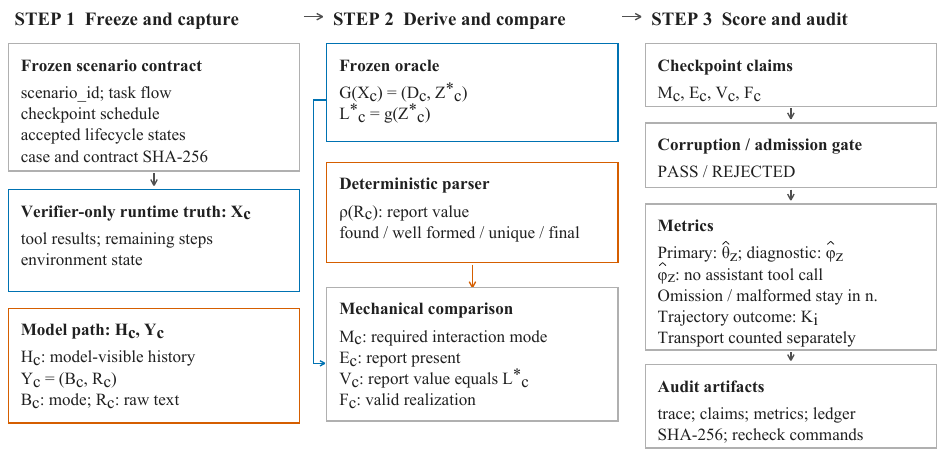}
\caption{StageIF's auditable evaluation pipeline.  Runtime truth is
derived independently of model answers.  Comparing it with parsed reports
yields checkpoint verdicts, failure codes, evidence references, and
recomputable metrics.}
\label{fig:auditable-pipeline}
\end{figure*}

Figure~\ref{fig:auditable-pipeline} summarizes the full StageIF evidence
path.  It is a process view, not an empirical result: no rates or model
comparisons are introduced by the figure.  The two evidence lanes are
kept distinct because model-visible history and stored model output do
not have authority over verifier-only runtime truth.  The frozen oracle
derives applicability and lifecycle truth, the deterministic parser
extracts the submitted report, and only then does a mechanical comparison
produce checkpoint claims, failure codes, and aggregate estimators.

\section{Prompts and Protocol Text}
\label{app:prompts}

\begin{figure}[h]
\centering
\includegraphics[width=\columnwidth]{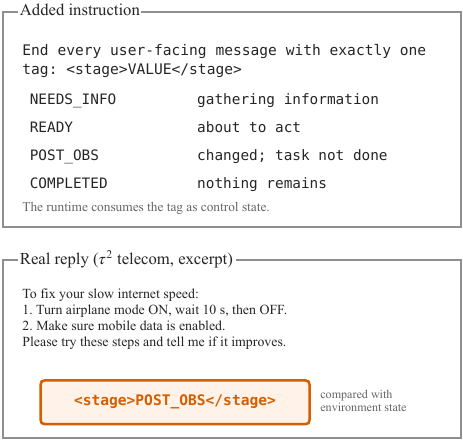}
\caption{The complete $\tau^2$-bench intervention.  The added
instruction defines four lifecycle values, and a verbatim response ends
with the machine-consumed tag.}
\label{fig:task-example}
\end{figure}

Frozen-checkpoint stimuli use a system prompt followed by a scripted
history; validating rollouts instead build subsequent history from
environment events.  The primary frozen scenarios are Chinese; the breadth
replication uses English scenarios whose system prompt is a
sentence-by-sentence rendering of the Chinese template.
Listing~\ref{lst:prompt-en} reproduces the English
system prompt verbatim (scenario \texttt{en-sched-01}, XML encoding,
baseline arm).  The Chinese template has the same structure
(role, two tools, clarify-then-confirm rule, output protocol); it cannot
be typeset in this template, so the remaining prompts are given as
faithful English renderings with the SHA-256 prefix of the original
string, which is stored verbatim in the artifact.  The exact generator and frozen stimulus of every prompt family are
indexed in the artifact repository (Appendix~\ref{app:repro}).

\begin{lstlisting}[style=stageif,caption={English system prompt, verbatim (breadth replication).},label={lst:prompt-en}]
You are a scheduling assistant. You can call the tools create_event and send_invites to fulfil the user's request. Ask the user for clarification when required information is missing; confirm with the user before performing any consequential operation. Output protocol: whenever you produce a natural-language reply based on a tool result, you must append the current task-stage marker at the very end of the reply, one of: <stage>needs_information</stage>, <stage>ready_for_action</stage>, <stage>post_observation_continue</stage>, <stage>completed</stage>. Meanings: needs_information = required information is missing and the user must clarify; ready_for_action = information is complete, awaiting user confirmation; post_observation_continue = an execution result exists but the task is not finished; completed = the task is finished.  Tool-call turns must not emit the marker; the marker must be unique and placed at the very end.
\end{lstlisting}

\paragraph{Protocol encodings.}  The JSON encoding replaces each
\texttt{<stage>...</stage>} example with \texttt{\{"stage": "..."\}};
both encodings use the same four values and the same placement rule.
The parser accepts a marker only if it is well formed, unique, and the
last non-empty line of the reply.

\paragraph{Arm variants (renderings of the Chinese originals).}
\begin{itemize}\setlength\itemsep{0pt}
\item \textbf{Static-tag arm} (\texttt{7ee8a3f1d989206b}): ``Output protocol: whenever you produce a natural-language reply based on a tool result, you must append the fixed protocol marker \texttt{<stage>active</stage>} at the very end (the value never changes).  Tool-call turns must not emit the marker.''
\item \textbf{Inert-tag arm} (\texttt{a4fc054ff443441d}): same rule with the semantically inert constant \texttt{<stage>qzx</stage>}, described as ``a meaningless fixed code that never changes.''
\item \textbf{Gold-injected arm}: the Chinese original is generated from a
fixed template whose faithful English rendering is ``[State hint] Current
task stage: \textit{value}.''  The original \textit{value} is exactly
\texttt{action}, \texttt{needs\_information},
\texttt{ready\_for\_action}, \texttt{post\_observation\_continue}, or
\texttt{completed}.  The corresponding original-string SHA-256 prefixes
are \texttt{47cdbf2810251959}, \texttt{97e72c4774a3ca71},
\texttt{af3b8051a3f7bc82}, \texttt{b977f3d9a8ade950}, and
\texttt{4239c8490fb5b24b}.
\item \textbf{Trailing reminder} (\texttt{deeeb876ddfb8c94}): an additional system message ``[Reminder] If you reply in natural language this turn, you must append the required stage marker at the very end; tool-call turns must not append it.'', appended as the last message of the frozen history, so the model sees it immediately before it answers, in the same request.
\end{itemize}

\paragraph{Matched obligations (Appendix~\ref{app:bridge}).}
Deferred terminal duty (\texttt{ef55f1c69df542c4}): ``Output protocol: when the entire task is complete, your final natural-language reply must end with \texttt{<sig>done</sig>}; no reply before completion may emit this marker.''  Stage-triggered status sentence (\texttt{a09baf03265cca3e}): ``Output protocol: whenever you produce a natural-language reply based on a tool result, if the task is not yet complete the reply must end with the sentence `(The task is still in progress.)'; if the task is complete it must end with `(The task is complete.)'.  Tool-call turns must not emit either sentence.''  The stage-triggered machine report uses the standard protocol above.

\paragraph{Continuation-context factors (\S\ref{sec:affordance}).}
The fully crossed follow-up varies three fields in the tool result.  Two
English wording families are used verbatim: ownership
\texttt{this\_assistant}/\texttt{another\_system} or
\texttt{assigned\_to\_you}/\texttt{assigned\_to\_external\_service};
cue \texttt{prior\_step\_succeeded}/\texttt{prior\_step\_returned\_an\_error}
or \texttt{no\_faults\_reported}/\texttt{fault\_reported\_upstream};
the third family renders the same pairs in Chinese.  Executability is
varied by declaring or withdrawing the second-step tool.

\paragraph{Validating rollout (Appendix~\ref{app:gate}).}
The runtime-owned arm re-prompts the model with a user-level message
(\texttt{7ce5a576378a52a9}): ``[System] The task is not yet complete.  Proceed with the information you have, use defaults for the rest, and execute the remaining steps directly.''  The message-matched gates B1 and B2 receive identical user messages and differ only in the termination rule.

\section{Framework termination-signal survey}
\label{app:survey}

\paragraph{How the survey was conducted.}  Seventeen widely used
agent frameworks were selected purposively for ecosystem coverage, not sampled, and each was asked one question: \emph{how does the agent
loop decide whether to continue or stop after this turn, and where does
the signal that decision reads come from?}  Answers were classified
into three families fixed before collection.
\textbf{(A)~Runtime-owned}, where the decision is computed by framework
code from structured state, covers graph edges and typed state (A1;
\citealp{fw:llamaindex,fw:dify,fw:langgraph}) and a hard
iteration or budget cap (A2;
\citealp{fw:semantickernel,fw:claudesdk,fw:googleadk,fw:dify,fw:metagpt,fw:crewai}).
\textbf{(B)~Model-emitted}, where the decision reads a signal the model
generated, covers an explicit textual marker (B1, the family isomorphic
to our setting; \citealp{fw:camel,fw:autogen}), the structural presence
or absence of tool calls (B2;
\citealp{fw:openaiagents,fw:claudesdk,fw:langgraph,fw:langchain,fw:vercelai,fw:agno}),
and a dedicated terminal tool (B3;
\citealp{fw:smolagents,fw:googleadk,fw:autogpt}); one framework is
model-emitted with its sub-type undocumented
(\citealp{fw:mastra}).  \textbf{(C)~Hybrid} covers frameworks that
expose more than one of these at once
(\citealp{fw:langgraph,fw:googleadk}).  B2 and B3 count as model-emitted because what the
runtime consumes is still a generated decision.  Collection and
adjudication were separated, every entry carries a primary source, a
verbatim quotation, and an access date, and a missing field forces
\textsc{evidence\_incomplete} rather than an inferred classification;
the collection protocol and per-framework evidence are in the artifact
repository (Appendix~\ref{app:repro}).

\begin{table*}[t]
\centering\scriptsize
\caption{Representative lifecycle-control interfaces, grouped by the
signal consumed by the runtime rather than ranked. Rows may overlap and
are not prevalence counts.}
\label{tab:framework-signal-map}
\setlength{\tabcolsep}{3pt}
\begin{tabular}{p{0.08\textwidth}p{0.19\textwidth}p{0.27\textwidth}p{0.13\textwidth}p{0.25\textwidth}}
\toprule
Family & Runtime consumes & Representative framework forms & Immediate authority & Relation to StageIF \\
\midrule
B1: textual marker & marker in generated text & CAMEL marker; AutoGen \texttt{TextMentionTermination} \citep{fw:camel,fw:autogen} & model text & Directly isomorphic to the studied channel. \\
B2: tool-call structure & presence of tool calls & OpenAI Agents, Claude Agent SDK, and LangGraph prebuilt ReAct rules \citep{fw:openaiagents,fw:claudesdk,fw:langgraph} & model structure & Different encoding; model output still controls continuation. \\
B3: terminal tool & dedicated completion call & smolagents \texttt{final\_answer}; Google ADK termination tools \citep{fw:smolagents,fw:googleadk} & model action & Structured model-issued completion decision. \\
A1: typed runtime state & event, edge, or expression & LlamaIndex \texttt{StopEvent}; Dify condition; LangGraph primitive edges \citep{fw:llamaindex,fw:dify,fw:langgraph} & runtime logic & Continuation derives from explicit system state. \\
A2: hard bound & iteration, turn, or budget counter & Caps in Semantic Kernel, Claude Agent SDK, Google ADK, Dify, MetaGPT, and CrewAI \citep{fw:semantickernel,fw:claudesdk,fw:googleadk,fw:dify,fw:metagpt,fw:crewai} & runtime counter & Backstop, not evidence of task completion. \\
Hybrid & multiple signals & model-issued termination plus runtime rules & mixed & Separates normal completion from failure containment. \\
\bottomrule
\end{tabular}
\end{table*}

\paragraph{Classification.}
Table~\ref{tab:framework-signal-map} normalizes heterogeneous interfaces
into the pre-specified families. In canonical usage, 13/17 frameworks
use model-emitted signals, 3/17 are runtime-owned, and 1/17 remains
evidence-incomplete. LangGraph exposes both A1 primitives
and B2 prebuilt routing; Google ADK combines B3 tools with an optional A2
cap. MetaGPT's model signal subtype is undocumented. CrewAI documents
\texttt{max\_iter} but not normal completion, so it remains
\textsc{evidence\_incomplete} rather than imputed. Semantic Kernel's
concrete default cap is 5; its abstract base exposes 99, which is not
the concrete default.

\paragraph{Recorded termination criteria.}  Each classification above
rests on the recorded documentation excerpt.  OpenAI Agents SDK \citep{fw:openaiagents}:
``the rule for whether the LLM output is considered as a `final output'
is that it produces text output with the desired type, and there are no
tool calls'' (B2).  Claude Agent SDK \citep{fw:claudesdk}: ``turns
continue until Claude produces output with no tool calls'' (B2, with a
\texttt{max\_turns} cap).  smolagents \citep{fw:smolagents}: ``in the end
you have to return a final answer using the \texttt{final\_answer} tool''
(B3).  Google ADK \citep{fw:googleadk}: the deprecated \texttt{LoopAgent}
used an \texttt{exit\_loop} tool and an optional \texttt{max\_iterations}
cap, while its successor \texttt{Workflow} exposes a dedicated
\texttt{finish\_task} tool (B3; the legacy cap is A2 when configured).
CAMEL \citep{fw:camel} matches our setting most directly: the loop breaks
on \texttt{if "CAMEL\_TASK\_DONE" in user\_response.msg.content} (B1).
AutoGen \citep{fw:autogen} ships pluggable strategies, of which the
keyword-based \texttt{TextMentionTermination} is the one used in
canonical tutorials (B1).  LlamaIndex Workflows \citep{fw:llamaindex}: ``when the
workflow encounters a returned \texttt{StopEvent}, it immediately stops''
(A1).  MetaGPT \citep{fw:metagpt}: a loop that runs ``until the role
thinks it is time to stop,'' with an iteration cap in parallel (B$+$A2,
sub-type not documented).  The remaining four records complete the
seventeen.  LangGraph \citep{fw:langgraph} is recorded under both
readings, since its graph primitives expose
\texttt{add\_conditional\_edges} (A1) while its prebuilt ReAct executor
stops on the absence of tool calls (B2).  Semantic Kernel
\citep{fw:semantickernel}: ``maximum\_iterations: int = Field(default=5,
description=\dots)'' in the concrete strategy, against 99 in the
abstract base, which is not the concrete default (A2).  Dify
\citep{fw:dify}: ``The loop terminates when either the termination
condition is met, the maximum count is reached, or an Exit Loop node
executes'' (A1$+$A2; this excerpt was retrieved on 2026-08-24, later
than the other documentation records).  CrewAI \citep{fw:crewai}
documents \texttt{max\_iter} but no criterion for normal completion, so
it is recorded as \textsc{evidence\_incomplete} rather than imputed; a
source-level reading resolves it, and we do not carry that reading into
the count.

The five frameworks added on 2026-09-02 follow.  LangChain's current
\texttt{create\_agent} API \citep{fw:langchain}: ``the process repeats
until no more \texttt{tool\_calls} are present in the response'' (B2).
AutoGPT's classic loop \citep{fw:autogpt}: the model calls a
\texttt{finish} tool, raising \texttt{AgentTerminated} (B3); the
project's newer visual-builder product line documents deployment but
not execution control, so only the classic loop is classified here.
Vercel AI SDK \citep{fw:vercelai}: ``a finish reasoning other than
tool-calls is returned\ldots or a stop condition is met,'' with a
default \texttt{stepCountIs(20)} cap (B2 with an A2 backstop).  Mastra
\citep{fw:mastra}: the loop ``continue[s] iterating until the model
emits a final answer or an optional stop condition is met'' (model-emitted,
sub-type not documented, with an optional A2 cap).  Agno
\citep{fw:agno}: ``No tool calls or finished processing them: break''
(B2).  All five consume a model-generated signal; none is
runtime-owned.

\paragraph{What the count does and does not support.}  Within this
sample, delegating the continue/stop decision to a model-generated
signal is the majority pattern, and the strictly isomorphic form---an
explicit textual marker---is default or tutorial-canonical in two
frameworks with a third offering it optionally.  The survey supports
prevalence within seventeen purposively chosen frameworks at two points
in time; it does not support universality, and framework defaults change.
We accordingly read StageIF as a stress test of whether a
model-generated lifecycle signal is fit to serve as control authority,
not as a claim that the four-value textual protocol itself is widely
deployed---what the survey establishes is that the continue/stop
decision is routinely delegated to some model-emitted signal.  The B2
family matters beyond marker-based designs for a structural
reason: when a loop stops because a turn contained no tool call, every
natural-language turn is implicitly emitting a continue/stop signal, so
the divergence failure we measure---acting where the stage called for
speaking---is in those runtimes a wrong lifecycle signal rather than a
missing one.  Appendix~\ref{app:gate} measures that concern in a rollout:
under a no-tool-call gate, 45.5\% of trajectories stall because the
agent spoke at a point that called for action, handing control to a
user with nothing to add.

\begin{sloppypar}
\paragraph{Source-level extension: five deployed runtimes.}  The survey
above is documentation-first and pins no commit.  We re-asked its single
question of five runtimes that ship inside deployed products---voice,
home automation, coding agents---rather than orchestration SDKs, this
time reading the source at a pinned commit.  Each record carries a
repository, that commit, a \texttt{file:line}, and a verbatim quotation;
access dates are 2026-08-18/19.  This adds coverage in a different
ecological niche.  It does not close the three evidence gaps above,
which remain as reported.

\textbf{Home Assistant} \citep{fw:homeassistant} pairs a cap with a
structural test: \texttt{MAX\_TOOL\_ITERATIONS = 10}
(\path{components/anthropic/entity.py:145}) around \texttt{if not
chat\_log.}\allowbreak\texttt{unresponded\_tool\_results: break} (\texttt{:1250}), the
property being whether the last message is a tool result
(\path{components/conversation/chat_log.py:377}).  The same shape
appears in all nine of its LLM conversation integrations (C, A2$+$B2).
\textbf{Pipecat} \citep{fw:pipecat} branches on the structured field
\texttt{chunk.choices[0].delta.tool\_calls}
(\path{services/openai/base_llm.py:509}); the continuation is gated on
\texttt{if frame.result:}
(\path{aggregators/llm_response_universal.py:1814}), so a tool
returning a falsy result triggers no further inference (B2).
\textbf{LiveKit Agents} \citep{fw:livekitagents} sets
\texttt{reply\_required = fnc\_out is not None}
(\path{voice/generation.py:1038})---a tool returning nothing requires
no reply---and on reaching \texttt{max\_tool\_steps} forces
\texttt{tool\_choice="none"} to avoid stopping silently
(\path{voice/agent_activity.py:3554}) (C, B2$+$A2).
\textbf{Pydantic AI} \citep{fw:pydanticai} routes on \texttt{if
tool\_calls:} in the node whose docstring reads ``decides whether to end
the run or make a new request''
(\path{pydantic_ai/_agent_graph.py:1817,2017}) (B2).
\textbf{OpenHands} \citep{fw:openhandssdk} caps at
\texttt{max\_iteration\_per\_run = 500}
(\path{sdk/conversation/conversation.py:75}) over a text-versus-tool
branch (C, B2$+$A2).

All five read a structural tool-call signal rather than a textual
marker, so none is exposed to the dead-end mode that only a marker-reading
loop can have.  None of the five, however, makes an intermediate report
a condition of continuing: progress surfaces through separate structured
channels the model need not write to---Home Assistant streams tool
deltas to the frontend, Pipecat and LiveKit emit function-call lifecycle
frames, and LiveKit's mid-tool update is opt-in.  Pydantic AI's deferred
tool requests are the one enforced third outcome, and they cover
external approval rather than progress.

One record is worth stating on its own because it is a naming fact
rather than a rate.  In OpenHands, a model reply carrying text and no
tool call sets \path{ConversationExecutionStatus.FINISHED}
(\path{sdk/agent/response_dispatch.py:250}, in a method documented as
``Handle LLM response with text content --- finishes conversation''),
while the same enumeration defines \texttt{FINISHED} as ``completed the
current task'' and a separate \texttt{STUCK} as ``stuck in a loop or
unable to proceed'' (\path{sdk/conversation/state.py:57,59}).
\texttt{STUCK} is live elsewhere in the same runtime, assigned by a
dedicated detector (\path{conversation/impl/local_conversation.py:726}).
A run that goes quiet mid-task is therefore recorded under the status
reserved for completion, though a status for the other reading exists
and is used.
\end{sloppypar}

\section{$\tau^2$-bench stage and position denominators}
\label{app:tau2-deciles}

Table~\ref{tab:tau2-stage} gives the counts behind
Figure~\ref{fig:tau2-stage-all} and the
four-deployment mean of \S\ref{sec:tau2} for six of the seven
compliant full-lane deployments under the strict parser, plus the
envelope-stripped rows behind the four-generation comparison in
\S\ref{sec:tau2}; \texttt{gpt-5.6-sol}'s malformed
tags count as wrong in the strict rows.  The envelope-stripped reading of that table
parses the reply as JSON, takes its \texttt{message} field (or the
bare string when the whole reply is a JSON string literal), and applies
the unchanged parser to it; replies of any other JSON shape carry no
readable tag and score as omitted reports in the stripped rows (12 of
\texttt{gpt-5.6-sol}'s 13 omissions, \texttt{gpt-6-astra}'s 1).
The three-way completion split in \S\ref{sec:tau2} separates the 20
hand-off tasks (reward basis includes the transfer action), then
classifies the remaining completion checkpoints by whether a customer
message between the trajectory's last non-\completed{} duty checkpoint
and the checkpoint carries a resolution cue without a negation (the
script's window, which can open before the completing tool call);
completions caused by the agent's own tool call on those tasks
(33--70\% correct) are counted with the unconfirmed group in the text's
ranges.  An independent replay that locates the first true completing
event from full tool responses moves a handful of checkpoints between
the agent-caused and unconfirmed groups for \texttt{claude-sonnet-5}
and \texttt{grok-4.6} (\texttt{claude-sonnet-5} 3/23 instead of 2/21
unconfirmed) and leaves
the hand-off and cue-confirmed counts unchanged.  The mid-task quintiles take
each trajectory's \postobs{} checkpoints, min--max-normalize their
turn indices, and cut the span into five; trajectories with a single
\postobs{} checkpoint have no depth and are omitted from the quintile
columns (23 checkpoints across the six strict rows; 2 and 6 in the two
stripped rows).  Gold rule: a checkpoint is \completed{} when the
task's environment assertions hold and, for tasks whose reward basis
also names a required action (the 20 transfer tasks of the base split),
that action has been called; an earlier rule that ignored the action
requirement scored those tasks complete from their first turn and was
replaced after an audit.  A completion checkpoint counts as
\emph{confirmed} when a customer message in that window contains a
resolution cue (works,
fixed, resolved, thanks, connected, and similar) and no negation
(still, not working, unable, and similar).  An earlier \texttt{gemini-3.7-flash} lane run at
vendor-default thinking is superseded by the \texttt{gemini-3.8-flash}
lane at minimal effort; it is retained in the artifact repository and
not reported here.  Rows are recomputed
from the scored lanes by the figure generators in the artifact
repository.

\paragraph{Per-deployment detail behind \S\ref{sec:tau2}.}
Among the deployments that largely ignore the duty, the highest reports
on 10.1\% of duty-active checkpoints (165/1640).  The two compliant
deployments that return no reasoning tokens report correctly at
90.6--99.4\% of pre-action checkpoints, and 94.0--99.6\% of their
mid-task errors, together with \texttt{claude-sonnet-5}'s, carry a
well-formed but wrong value rather than no value.  Across the four
generations of one family, ordered by release, the mid-task rate moves
from 5.8\% to 11.5\%, 42.3\%, and 100.0\% and the completion rate from
88.9\% to 87.3\%, 67.5\%, and 48.1\%, the last two generations read
after envelope stripping; an eight-task probe of
\texttt{claude-fable-5.1} against \texttt{claude-sonnet-5} gives 78\%
versus 6\% mid-task at probe scale only.  The JSON envelope covers
34.7\% (332/958) of \texttt{gpt-5.6-sol}'s duty checkpoints and 100\%
(1{,}091/1{,}091) of \texttt{gpt-6-astra}'s, against 1 of 1{,}023
replies for \texttt{gemini-3.8-flash} and 0 of 134 for a
\texttt{claude-fable-5.1} probe; stripping it moves
\texttt{gpt-5.6-sol}'s mid-task rate from 28.6\% to 42.3\% and
\texttt{gpt-6-astra}'s picture from 0\% throughout to 99.6\% before
acting, 100\% mid-task, and 48.1\% at completion.
\texttt{gemini-3.8-flash}'s mid-task errors are 78\% stale ($n{=}183$).
The three deployments that stay accurate mid-task hold, in order,
74.1\% mid-task ($n{=}707$) and 63.8\% at completion ($n{=}94$) for
\texttt{gemini-3.8-flash}; 100\% ($n{=}726$) and 48.1\% ($n{=}135$) for
\texttt{gpt-6-astra} read after envelope stripping; and 94.7\%
($n{=}780$) and 63.6\% ($n{=}129$) for \texttt{claude-opus-5} on its
106 single-simulator tasks.
In the benchmark's no-user mode, ending the episode on the model's own
\texttt{done()} signal scores zero on 63 of 114 tasks for \texttt{gpt-4.1} and 30
of 114 for \texttt{gpt-5.5}.
The native stop signal is not the added stage tag, and zero reward
alone does not establish a false completion claim.

Within the mid-task stage, accuracy falls from the first to the last
quintile: 30.5\% $\rightarrow$ 5.1\%, 21.7\% $\rightarrow$ 7.3\%, and
12.4\% $\rightarrow$ 4.7\% for \texttt{claude-sonnet-5},
\texttt{gpt-5.5}, and \texttt{gpt-4.1}, with trajectories of 15 or more
duty checkpoints worse still; the two envelope-stripped generations
hold at 49.4\% $\rightarrow$ 47.1\% and at 100\% throughout.  Errors
cluster by trajectory: among trajectories with two or more mid-task
checkpoints, every mid-task report is wrong in 67\% (\texttt{gpt-4.1},
73/109), 67\% (\texttt{gpt-5.5}, 64/95), and 44\%
(\texttt{claude-sonnet-5}, 43/98) of them, while
\texttt{gemini-3.8-flash} is 41\% entirely correct and 18\% entirely
wrong and \texttt{gpt-6-astra} entirely correct throughout.  Lost
trajectories concentrate in the multi-fault MMS task family (39/49,
41/48, and 28/45 for the three collapsing deployments) rather than in
service tasks (13/25, 6/16, and 5/18).  Post-transfer on the hand-off
tasks, the newest deployments report \completed{} on 0--40\% of
checkpoints (\texttt{gpt-6-astra} 0/18, \texttt{gpt-5.6-sol} 5/20,
\texttt{gemini-3.8-flash} 8/20).  Before a resolution cue,
and outside the hand-off tasks, \texttt{gpt-6-astra} reports
\completed{} on 0/42 such checkpoints, \texttt{gemini-3.8-flash} on
0/19, and \texttt{claude-sonnet-5} on 3/29; the remaining deployments
have 3 to 34 such checkpoints.

\paragraph{Pooled mean and thinking status per lane.}  The four
compliant deployments run without requesting thinking
(\texttt{claude-sonnet-5}, \texttt{gpt-4.1}, \texttt{gpt-5.5}, and
\texttt{gpt-5.6-sol}, whose malformed tags count as wrong) average
83.7\% before acting, 11.1\% at the lowest mid-task quintile, and
75.1\% at completion on the stage axis.  The controlled testbed does
not reproduce \texttt{claude-opus-5}'s terminal signature: there it
scores 98.8\% at the finished checkpoint ($n{=}240$;
\S\ref{sec:results}), so that shape is setting-specific rather than a
fixed property of the deployment.  The four pooled deployments were
run without requesting thinking;
\texttt{gpt-4.1} and \texttt{gpt-5.5} returned no reasoning tokens,
but \texttt{claude-sonnet-5} and \texttt{gpt-5.6-sol} returned them
on 64\% and 82\% of calls (median 68 and 56 tokens when present), so
that mean is ``thinking not requested,'' not ``thinking absent.''
\texttt{grok-4.6}, whose extended thinking cannot be switched off
(reasoning tokens on 99\% of calls, median 332), falls from 96.4\%
before acting to 45.5--57.9\% across the mid-task quintiles ($n{=}281$
and 96--181 per quintile).  \texttt{gemini-3.8-flash} and
\texttt{gpt-6-astra} refuse to disable reasoning and run at the lowest
effort their APIs allow (median 0 reasoning tokens; 7.2\% and 1.0\% of
calls nonzero).  The \texttt{claude-opus-5} lane ran with thinking
off; a gateway quota interruption changed its user-simulator model
after 8 of 114 trajectories, and the remaining 106, run on the
replacement simulator that the newest-generation lanes also use, are
the ones reported (mid-task 94.7\%, $n{=}780$; completion 63.6\%,
$n{=}129$; mid-task 94.2--95.3\% in each simulator and cache stratum).
A 40-task run of the same model with vendor-default thinking on the
original simulator (mid-task 89.0\%, $n{=}391$) is retained in the
artifact repository and not reported here.

\begin{figure}[h]
\centering
\includegraphics[width=\columnwidth,keepaspectratio]{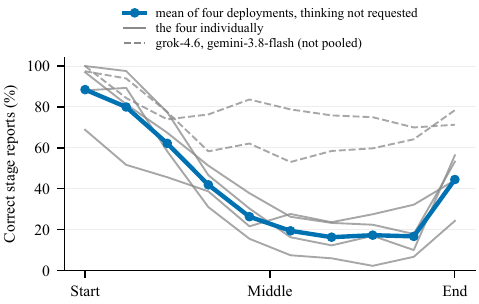}
\caption{Position-normalized view of the telecom data.  The horizontal
axis is the share of duty checkpoints elapsed, cut into ten
within-trajectory bins with lifecycle states mixed.}
\label{fig:teaser-tau2-all}
\end{figure}

Figure~\ref{fig:teaser-tau2-all} shows the same data on a
position-normalized axis instead: each trajectory's duty-active
checkpoints are pooled into ten min--max-normalized position bins
regardless of state, the axis of Lost-in-the-Middle-style plots.  This
view mixes lifecycle states within a bin---the last bin holds both
completion checkpoints and mid-task checkpoints of trajectories the
user ended early---which is why the paper's main figure is drawn by
state.

\begin{table*}[h]
\centering
\small
\setlength{\tabcolsep}{5pt}
\begin{tabular}{@{}lrrrrrrr@{}}
\toprule
Deployment & Before acting & Mid Q1 & Mid Q2 & Mid Q3 & Mid Q4 & Mid Q5 & Completed \\
\midrule
\texttt{claude-sonnet-5} & 410/516 & 54/177 & 21/123 & 15/139 & 19/122 & 9/175 & 190/226 \\
\texttt{gpt-4.1} & 393/434 & 25/202 & 3/136 & 5/114 & 2/123 & 8/170 & 32/36 \\
\texttt{gpt-5.5} & 469/472 & 34/157 & 10/111 & 8/105 & 9/92 & 12/165 & 96/110 \\
\texttt{gpt-5.6-sol} & 160/245 & 52/160 & 22/109 & 23/107 & 30/103 & 55/155 & 31/77 \\
\midrule
\texttt{grok-4.6} & 271/281 & 101/181 & 45/99 & 55/100 & 52/96 & 84/145 & 130/140 \\
\texttt{gemini-3.8-flash} & 219/222 & 121/180 & 91/117 & 86/114 & 90/118 & 133/171 & 60/94 \\
\midrule
\texttt{gpt-5.6-sol}, envelope stripped & 234/245 & 79/160 & 39/109 & 33/107 & 45/103 & 73/155 & 52/77 \\
\texttt{gpt-6-astra}, envelope stripped & 229/230 & 173/173 & 108/108 & 120/120 & 132/132 & 187/187 & 65/135 \\
\bottomrule
\end{tabular}
\caption{Correct/total stage reports at each state-axis point (before
acting, five mid-task quintiles, completed) for six compliant full-lane
telecom deployments under the strict parser; \texttt{gpt-5.6-sol}'s
malformed tags count as wrong.  The last two rows give the
envelope-stripped readings of the two JSON-wrapping deployments.  Rows
are grouped, not ranked.}
\label{tab:tau2-stage}
\end{table*}

\section{Per-experiment admission and exclusion}
\label{app:admission}

\paragraph{Measurement discipline.}  The 50{,}400 logical positions
are a repeated-measures count over 12 scenario clusters with 20
repetitions, not 50{,}400 independent tasks.  A corruption gate that
must collapse every success metric on blanked and value-inverted
outputs passed on live data (Appendix~\ref{app:verifier}); transport
and identity failures are reported on their own denominators
(Appendix~\ref{app:limits}).  Behavioral results, the framework
documentation audit (Appendix~\ref{app:survey}), and the source audit
of deployed runtimes are never pooled.

Each study froze its own identity precheck, so the admitted deployment set
differs across studies.  Table~\ref{tab:admission} lists every planned deployment,
the admitted count, and the reason for each exclusion, reconstructed from the
archived \texttt{prechecks.json} and \texttt{summary.json} of each run.  No
excluded deployment consumed any denominator; exclusions are reported, never
imputed.  The nine planned deployments are deepseek-v4-flash,
deepseek-v4-pro, glm-5.1, gpt-4.1, gpt-4.1-mini, gpt-5.5,
qwen3.5-122b-a10b, qwen3.5-35b-a3b, and qwen3.6-27b, each with a frozen generation configuration and
listed alphabetically throughout; the identifiers name these bundles,
and no ordering in this paper is a ranking.

Every planned deployment is marketed by its vendor for agentic or
tool-use work.  We checked each vendor's own model card, release post, or
API documentation and found an explicit claim for eight of the nine,
ranging from a one-line positioning statement to a dedicated section on
long-horizon execution; the claims differ in prominence, not in
presence, so this sample supports no contrast between agent-oriented and
other deployments.  Two caveats attach to the check: the two
\texttt{deepseek-v4-*} identifiers are rolling API aliases whose
weight snapshot at run time cannot be recovered from the identifier
alone, and two vendor blogs were readable only through web archives at
the time of writing.  This is a documentation observation about vendor
positioning, not an evaluation of the capabilities claimed.

\begin{table*}[t]
\centering\scriptsize
\caption{Admission per study.  ``Planned'' is the frozen deployment list; exclusions
name the deployment and the mechanical reason.  The repair diagnostics replay a
deployment's own prior reply, so their populations are sampled from the
deployments that produced those replies rather than from the full deployment list.}
\label{tab:admission}
\setlength{\tabcolsep}{4pt}
\begin{tabular}{p{0.23\textwidth}rrp{0.45\textwidth}}
\toprule
Study (section) & Planned & Admitted & Excluded and reason \\
\midrule
Stage pattern (\S\ref{sec:rq1}) & 9 & 7 & gpt-4.1, gpt-4.1-mini: identity precheck failed (repeated transport failures) \\
Five-arm contrasts (\S\ref{sec:factorial}, App.~\ref{app:stats}) & 9 & 7 & same two deployments, same reason \\
Matched-obligation bridge (App.~\ref{app:bridge}) & 9 & 9 & --- \\
Negative controls (App.~\ref{app:negatives}) & 9 & 9 & --- \\
Breadth replications (App.~\ref{app:breadth}) & 9 & 9 & --- \\
Protocol-demand rerun (App.~\ref{app:metrics}) & 9 & 9 & --- \\
Thinking-mode A/B (App.~\ref{app:thinking}) & toggle-exposing deployments & 4 verified & two endpoints inline reasoning into the answer channel; gpt-5.5 manipulation unverifiable \\
Withdraw-and-delegate bundle (\S\ref{sec:affordance}) & 9 & 9 & --- \\
Fully crossed decomposition (\S\ref{sec:affordance}) & 9 & 8 & gpt-4.1-mini: frozen precheck failed \\
Validating rollout, all gates (Appendix~\ref{app:gate}) & 9 & 8 & gpt-5.5: serving-side outage during the batch (3/20 probe successes, while another deployment in the same window succeeded on 90\%); backfill not attempted below the frozen 18/20 readiness bar \\
Replay and isolated question (App.~\ref{app:repair}) & 3 & 3 & populations sampled from the three deployments whose omitted-tag replies were replayed \\
Wrong-state follow-up (App.~\ref{app:repair}) & 9 & 6 & gpt-5.5, glm-5.1: lane completion 92.5\% and 94.3\%, below the pre-registered 95\% gate (infrastructure-missing); gpt-4.1-mini: no data after 25 precheck attempts across two runs \\
\bottomrule
\end{tabular}
\end{table*}

\section{Deployments and identifiers}
\label{app:models}

Nineteen model deployments appear in this paper.  The identifier is the
string the provider API exposed at run time; it names a
model-and-configuration bundle, not an entry in a ranking, and the
table is alphabetical.  The last column gives a public
page on which we confirmed that exact string, checked as a literal
match with word boundaries so that a longer dated variant cannot pass
for the identifier itself.  Identifiers are printed in the form used at
call time; where the vendor's own string differs only in punctuation or
case we say so in the row, One identifier is published only on its
vendor's Chinese-locale pages, which is where we confirmed it.  Three caveats carry over from the vendor check reported above.
The two \texttt{deepseek-v4-*} names are rolling aliases whose weight
snapshot at run time cannot be recovered from the identifier alone.
The \texttt{qwen3.5} identifiers match open-weight repository names,
whose correspondence to a hosted name we did not verify.  And the five
\texttt{gpt-*} identifiers are confirmed against the vendor's official
client library, whose model enumeration is generated from its API
specification, because its documentation pages were not retrievable
from our network.

\begin{table*}[t]
\centering\scriptsize
\caption{Model deployments used in this paper, alphabetical and not
ranked.  ``Where used'' names the studies; a deployment may serve in
one study and be excluded from another by that study's own admission
rule.  ``As run'' records the reasoning configuration actually
requested and, where measured, what the endpoint returned.  An empty
record cell means no public page carrying that identifier was located.}
\label{tab:models}
\setlength{\tabcolsep}{4pt}
\begin{tabular}{@{}p{0.16\textwidth}p{0.29\textwidth}p{0.31\textwidth}p{0.14\textwidth}@{}}
\toprule
Identifier & Where used & As run & Public record \\
\midrule
\texttt{claude-fable-5.1} & $\tau^2$ telecom, eight-task probe only & Minimal effort & \citep{card:claudemodels}; vendor string \texttt{claude-fable-5-1} \\
\texttt{claude-opus-5} & $\tau^2$ telecom; StageIF later cohort & Reasoning off in $\tau^2$; vendor-default sampling in StageIF & \citep{card:claudemodels} \\
\texttt{claude-sonnet-5} & $\tau^2$ telecom & Thinking not requested; reasoning tokens returned on 64\% of calls & \citep{card:claudemodels} \\
\texttt{deepseek-v4-flash} & $\tau^2$ telecom, retail, no-user; StageIF; replacement user simulator & Reasoning off & \citep{card:deepseekv4flash,card:deepseekapi} \\
\texttt{deepseek-v4-pro} & $\tau^2$ telecom, retail, no-user; StageIF & Reasoning off & \citep{card:deepseekapi} \\
\texttt{gemini-3.7-flash} & $\tau^2$ telecom, superseded and not reported; StageIF later cohort & Vendor default & \citep{card:gemini} \\
\texttt{gemini-3.8-flash} & $\tau^2$ telecom & Lowest effort the API allows; reasoning tokens on 7.2\% of calls & \citep{card:gemini} \\
\texttt{glm-5.1} & $\tau^2$ telecom, retail; StageIF & Reasoning off & \citep{card:glm51} \\
\texttt{glm-5.3} & StageIF later cohort & Thinking on; cannot be disabled & \citep{card:glm53} \\
\texttt{gpt-4.1} & $\tau^2$ telecom, retail, no-user; StageIF; original user simulator & No reasoning returned & \citep{card:gpt41} \\
\texttt{gpt-4.1-mini} & StageIF & Frozen generation configuration & \citep{card:openaisdk} \\
\texttt{gpt-5.5} & $\tau^2$ telecom, retail, no-user; StageIF & Reasoning effort none; no reasoning returned & \citep{card:gpt55} \\
\texttt{gpt-5.6-sol} & $\tau^2$ telecom & Thinking not requested; reasoning tokens returned on 82\% of calls & \citep{card:openaisdk} \\
\texttt{gpt-6-astra} & $\tau^2$ telecom & Minimal effort & \citep{card:openaisdk} \\
\texttt{grok-4.6} & $\tau^2$ telecom & Extended thinking on; cannot be disabled & \citep{card:xai} \\
\texttt{qwen3.5-122b-a10b} & StageIF & Frozen generation configuration & \citep{card:qwen35122b}; vendor string capitalised \\
\texttt{qwen3.5-35b-a3b} & StageIF & Frozen generation configuration & \citep{card:qwen3535b}; vendor string capitalised \\
\texttt{qwen3.6-27b} & StageIF & Frozen generation configuration & \citep{card:qwen3627b} \\
\texttt{qwen3.8-max} & $\tau^2$ telecom, among the deployments that usually omit the report & Reasoning off & \citep{card:modelstudio} \\
\bottomrule
\end{tabular}
\end{table*}

Two of the identifiers also served as the simulated customer rather
than as the agent under test.  \texttt{gpt-4.1} was the user simulator
for the earlier $\tau^2$ runs and \texttt{deepseek-v4-flash} for the
later ones, which is why the \texttt{claude-opus-5} lane spans both and
is reported on its single-simulator subset.

\section{Additional design and audit notes}
\label{app:limits}

\paragraph{Frozen policies and estimator conventions.}  Admission,
retry, transport, and endpoint-identity policies are frozen; complete
ledger and artifact identities are in this appendix and
Appendix~\ref{app:repro}.
All reported model-behavior rates instantiate the empirical estimators
of \S\ref{sec:construct} on $S_c=1$ checkpoints; transport and identity
failures are reported on their own denominators.  The oracle $G(X_c)$
and checkpoint membership are frozen before observing $Y_c$.

Endpoint exclusions, the first-turn applicability rule of the live
rollout, and the status of every designed follow-up are recorded in the
artifact repository (Appendix~\ref{app:repro}).

\section{Verifier-strictness forensics}
\label{app:verifier}

Under the initial verifiers the JSON parser demanded a standalone line
while the XML parser accepted line-end embedding.  A malformed-case
audit located this asymmetric strictness in the harness, not the
models; the stricter original verdicts are retained unmodified as a
forensic record, and equal-strictness rescoring surfaces 147 genuine
wrong-value cases, including hallucinated out-of-vocabulary stage
values (details in the artifact repository, Appendix~\ref{app:repro}).

\paragraph{Corruption gate.}  Replaying the full metric pipeline over
blanked and value-inverted outputs must collapse every success metric
(blanked: only $D_c{=}0$ rows may remain clean; inverted: zero)---any
metric surviving corruption measures the harness, not the model.
Deterministic scoring also sidesteps a documented failure of LLM judges
on this class of discrimination (AUROC at most 0.65, unimproved by
supplying the ground-truth specification) \citep{advani2026}.

\section{Motivating seed study}
\label{sec:seed}
\label{app:seed}

\paragraph{Role in this paper.}
An exploratory replay of the production protocol first revealed a
large terminal--intermediate asymmetry, dominated by systematic
omissions rather than malformed reports.  Because that study could not
separate stage, prompt, vocabulary, domain, or serving explanations, it
serves only as motivation; its design, denominators, and results are
reported below.  StageIF is the frozen instrument
used for the paper's primary evidence.

The seed protocol, reconstructed from locally desensitized motivating
artifacts, requires every post-tool natural-language turn to end with an
XML lifecycle tag: an intermediate value while awaiting user input and a
terminal value on completion.  We replayed four slot-filling
clarification scenarios at two independent frozen-history checkpoints
per trajectory over 22 endpoints across five deployment families.  All
requests used thinking off, temperature $0$, streaming, disabled SDK
retries, and successful identity prechecks.  The planned denominator was
$22\times4\times5=440$ trajectories.  The result archive is complete,
but the attempt ledger is not retained, so the historical total-request
count is not used as an auditable claim.

\begin{table}[h]
\centering\small
\caption{Motivating seed study on a fixed planned denominator, with
Wilson 95\% CIs.}
\label{tab:seed}
\begin{tabular}{lrr}
\toprule
Checkpoint & Valid/planned & 95\% CI \\
\midrule
Intermediate & 263/440 (59.8\%) & [55.1, 64.3] \\
Terminal     & 417/440 (94.8\%) & [92.3, 96.5] \\
\bottomrule
\end{tabular}
\end{table}

Per endpoint, intermediate validity is below terminal validity on 16/22,
tied on 3, and reversed on 3.  All
172 protocol failures are complete omissions; none is a wrong value,
duplicate, or malformed tag.  Fifteen intermediate checkpoints instead
issued a tool call, motivating the later separation of interaction-mode
divergence from report omission.  A 20-repetition extension on six
endpoints ($n=80$ each) exhibits three regimes: deterministic omission
(0/77 with zero variance), context-conditioned omission (19/19 in one
scenario and 0/59 in the others), and stochastic omission.  Thus the
seed establishes a motivating regularity, not its cause or external
scope.  Full endpoint tables, the three-regime analysis, and error-code
inventories are retained in the local evidence package.

\section{Bootstrap effect tables}
\label{app:stats}

\paragraph{Denominators and per-deployment detail behind \S\ref{sec:results} and \S\ref{sec:factorial-head}.}
Table~\ref{tab:gap} excludes transport failures and holds $n{=}478$--$480$
per cell; Figure~\ref{fig:teaser} plots $n{=}240$ planned per
deployment--checkpoint cell, the breadth replication $n{=}2160$ planned
per point, and the gold-injected and static-value arms $n{=}1680$
planned per arm--checkpoint point.  The later cohort reproduces the
stage shape with a smaller margin, 19.3 points for \texttt{glm-5.3} and
6.8 for \texttt{claude-opus-5} with all arms pooled, and
\texttt{gemini-3.7-flash} scores lowest one checkpoint later than the
other nine (87.8\% versus 69.9\%).  For two deployments the
intermediate decline is almost entirely mode selection: conditional
report validity falls only 1.8 and 5.6 points from terminal.  In the
five-arm batch, the trailing reminder cuts silent omission at the two
action-bearing checkpoints from 25.4\% to 16.6\% of checkpoints while
action divergence stays at 48.1\% against 49.7\% (per-deployment shifts
of $-7.1$ to $+12.3$ points); its largest gain is $+9.3$ points pooled
over all five checkpoints (CI $[+6.8,+11.8]$), or 11.7 points over
duty-active checkpoints alone.

\begin{table}[t]
\centering\scriptsize
\caption{\textsc{Controlled $2\times2\times2$ decomposition}.  Each
bracket is the range of deployment-specific contrasts, in percentage
points.  $\theta$ is primary; emission and false completion use the
fixed successful-response denominator and are diagnostic.  Tool
withdrawal is the one contrast whose adherence effect is non-negative on
all $8/8$ deployments; the other three are mixed in sign.}
\label{tab:factorial}
\setlength{\tabcolsep}{2pt}
\begin{tabular}{@{}p{0.29\columnwidth}rrr@{}}
\toprule
Contrast & $\Delta\widehat\theta$ & $\Delta$ emission & $\Delta$ false compl. \\
\midrule
Other $-$ self & $[-14.0,+9.0]$ & $[+4.4,+32.6]$ & $[+2.5,+44.1]$ \\
Withdrawn $-$ available & $[+0.3,+18.7]$ & $[6.8,56.4]$ & $[3.9,37.7]$ \\
Problem $-$ resolved & $[-4.7,+24.0]$ & $[-16.3,+12.3]$ & $[-13.1,+6.6]$ \\
Responsibility $\times$ executability & $[-30.6,+6.1]$ & $[-34.6,+23.8]$ & $[-7.5,+53.9]$ \\
\bottomrule
\end{tabular}
\end{table}

\begin{table}[t]
\centering\small
\caption{Intermediate vs.\ terminal checkpoints (all arms pooled):
the primary end-to-end metric $\widehat\theta$, diagnostic conditional
report validity $\widehat\phi$, report presence $E$, and correct withholding
$\widehat\omega$ at $D_c{=}0$.  $\widehat\phi$ conditions on no
assistant tool call; all columns exclude transport failures.}
\label{tab:gap-pooled}
\setlength{\tabcolsep}{2.5pt}
\resizebox{0.95\columnwidth}{!}{%
\begin{tabular}{lrrrrrr}
\toprule
Deployment & Int.\ $\widehat\theta$ & Term.\ $\widehat\theta$ & Int.\ $\widehat\phi$ & Term.\ $\widehat\phi$ & Int.\ $E$ & $\widehat\omega$ \\
\midrule
deepseek-v4-flash & 16.7\% & 64.0\% & 22.8\% & 64.0\% & 27.8\% & 100\% \\
deepseek-v4-pro & 24.1\% & 65.6\% & 32.5\% & 65.6\% & 34.5\% & 100\% \\
glm-5.1 & 17.3\% & 60.1\% & 23.6\% & 60.1\% & 28.7\% & 100\% \\
gpt-5.5 & 43.3\% & 58.4\% & 48.0\% & 58.4\% & 80.2\% & 100\% \\
qwen3.5-122b-a10b & 28.6\% & 61.3\% & 59.5\% & 61.3\% & 43.8\% & 98.8\% \\
qwen3.5-35b-a3b &  8.2\% & 97.4\% & 14.3\% & 97.7\% & 11.5\% & 100\% \\
qwen3.6-27b & 21.1\% & 54.6\% & 49.0\% & 54.6\% & 33.2\% & 100\% \\
\bottomrule
\end{tabular}
}
\end{table}

\begin{figure}[h]
\centering
\includegraphics[width=\columnwidth]{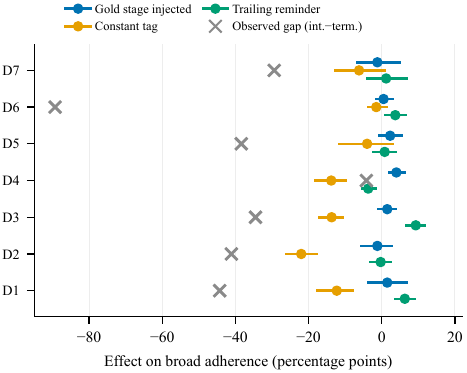}
\caption{Effects of three prompting interventions compared with each
deployment's observed intermediate--terminal gap.  No intervention
closes the gap.}
\label{fig:interventions}
\end{figure}

\paragraph{Five-arm factorial (\S\ref{sec:factorial}).}  600 stimuli
$\times$ 10 repetitions $\times$ 7 deployments $=$ 42{,}000
checkpoints, under the same measurement discipline as the pilot:
\textbf{A}~baseline (dynamic values, re-run as batch control);
\textbf{B}~constant semantic value; \textbf{C}~constant
\emph{semantically inert} value; \textbf{D}~trailing obligation
reminder $+$ constant value; \textbf{E}~reminder $+$ dynamic values
(reminder text and placement in Appendix~\ref{app:prompts}).
Baseline rates are directionally consistent with the pilot (e.g.\
46.7\% vs.\ 46.4\% on one deployment, same arm and denominator), an internal cross-batch check.
Figure~\ref{fig:interventions} plots every arm against each
deployment's observed gap.  Salience is E$-$A; semantics is C$-$B (little change on five
deployments, slight help on two); reminder$\times$constant is D$-$B
($+5$ to $+9$ on three deployments).  At the two action-bearing
checkpoints (CP2 and CP4; $n{=}3360$ per arm), E versus A shifts the
outcome mix from 25.4\% omission, 48.1\% divergence, 10.7\% malformed,
and 11.4\% clean to 16.6\%, 49.7\%, 14.4\%, and 14.0\%: the reminder
acts on the omission channel and leaves divergence unchanged.  The reminder that
\emph{reduces} compliance costs $-3.7$ points, CI $[-5.3,-1.7]$.
Figure~\ref{fig:interventions} uses clustered-bootstrap 95\% intervals;
crosses mark each deployment's observed intermediate--terminal gap.

\begin{table}[h]
\centering\small
\caption{Clustered-bootstrap effects on broad adherence with 95\% CIs;
$\dagger$ marks intervals that exclude zero.  E1: gold-injected minus
baseline.  E2: static minus baseline.  E3: intermediate minus terminal
in the baseline arm.}
\setlength{\tabcolsep}{3pt}
\begin{tabular}{lrrr}
\toprule
Deployment & E1 & E2 & E3 \\
\midrule
ds-v4-flash & $+1.5$ & $-12.3^{\dagger}$ & $-44.3^{\dagger}$ \\
ds-v4-pro & $-1.2$ & $-22.0^{\dagger}$ & $-41.1^{\dagger}$ \\
glm-5.1 & $+1.5$ & $-13.7^{\dagger}$ & $-34.5^{\dagger}$ \\
gpt-5.5 & $+4.0^{\dagger}$ & $-13.8^{\dagger}$ & $-4.2$ \\
qwen3.5-122b-a10b & $+2.3$ & $-4.0$ & $-38.4^{\dagger}$ \\
qwen3.5-35b-a3b & $+0.5$ & $-1.5$ & $-89.3^{\dagger}$ \\
qwen3.6-27b & $-1.2$ & $-6.2$ & $-29.4^{\dagger}$ \\
\bottomrule
\end{tabular}
\end{table}

\paragraph{Mixed-model robustness.}  The clustered bootstrap and the
exact permutation test both treat the scenario as the unit; a mixed
model is reported here as a third view rather than as a replacement.
Per deployment we fit \texttt{clean $\sim$ intermediate} with a
scenario random intercept as a linear mixed model, so the coefficient
is a probability difference comparable to the bootstrap estimates
(baseline arm, XML, $D_c{=}1$ rows; 6{,}718 observations).  Every
deployment is negative and significant: $-0.135$ (gpt-5.5,
$p{=}1.1\mathrm{e}{-}9$) through $-0.879$ (qwen3.5-35b-a3b,
$p{<}1\mathrm{e}{-}300$), reproducing the bootstrap ordering.  Pooled
across deployments as a logistic GEE with deployment fixed effects and
scenario clusters, the intermediate coefficient is $-5.435$
($\mathrm{se}\,0.823$, $p{=}4.0\mathrm{e}{-}11$; odds ratio $0.004$,
95\% CI $[0.001, 0.022]$).  Refitting with each scenario dropped in
turn moves the pooled coefficient only within $[-6.892, -5.296]$, so
no single scenario carries the effect.

\section{Diversified matched negative controls}
\label{app:negatives}

Three additional $D_c{=}0$ positions, each a minimal observable flip of
an existing checkpoint on the same twelve scenarios (gold derived by an
additive extension of the frozen state machine; pre-existing
derivations byte-identical under test): \emph{pre-tool} (the first
turn, before any tool event; flip of CP1), \emph{mid-chain} (user has
authorized the next step; correct turn is the second tool call; flip
of CP4), and \emph{post-done} (completion already reported, user says
thanks; the model must speak but the duty is off; flip of CP5).
Baseline arm, XML encoding, 20 repetitions, nine deployments; 6{,}480
checkpoints, zero terminal transport failures; the marker-injection
corruption gate collapses every clean verdict.

\begin{table}[h]
\centering\small
\caption{False-alarm rate at $D_c{=}0$ positions, where no protocol is
due; lower is better.  Post-conf.\ is the pilot's original tool-call
position on archived rows of the same arm and encoding; ``---'' marks a
deployment not admitted in the pilot.}
\label{tab:negatives}
\setlength{\tabcolsep}{3pt}
\resizebox{\columnwidth}{!}{%
\begin{tabular}{lrrrr}
\toprule
Deployment & Pre-tool & Post-conf. & Mid-chain & Post-done \\
\midrule
ds-v4-flash   & 54.2\%  & 0.0\% & 0.0\%  & 100.0\% \\
ds-v4-pro     & 86.7\%  & 0.0\% & 0.0\%  & 100.0\% \\
glm-5.1       & 98.8\%  & 0.0\% & 0.0\%  & 90.4\% \\
gpt-4.1       & 100.0\% & ---   & 0.0\%  & 100.0\% \\
gpt-4.1-mini  & 100.0\% & ---   & 0.0\%  & 63.3\% \\
gpt-5.5       & 100.0\% & 0.0\% & 25.2\% & 79.9\% \\
qwen3.5-122b-a10b& 49.6\%  & 1.2\% & 0.0\%  & 100.0\% \\
qwen3.5-35b-a3b  & 0.0\%   & 0.0\% & 0.0\%  & 100.0\% \\
qwen3.6-27b   & 41.7\%  & 0.0\% & 0.0\%  & 100.0\% \\
\bottomrule
\end{tabular}
}
\end{table}

Scenario-clustered bootstrap CIs are archived in the local evidence
package; the
post-done rates of exactly 100\% carry degenerate CIs
$[100\%,100\%]$ on six deployments.  A bare lifecycle tag as the entire reply occurs in 121 of the 2{,}000
post-done false alarms (6\%); the modal form is a one-line courtesy
followed by the tag.

\section{Breadth replications}
\label{app:breadth}

Six new it-helpdesk scenarios instantiate a second FSM shape---a
single consequential step with no mid-task stage
(clarify$\to$confirm$\to$act$\to$complete; checkpoints
\needsinfo/\ready/$D_c{=}0$/\completed)---and all twelve pilot
scenarios were translated to English (every surface string, including
tool argument and result strings; tools, FSM, and checkpoints
unchanged).  Baseline arm, XML encoding, 20 repetitions, nine
deployments; 15{,}120 checkpoints; corruption gates pass.

\begin{table}[h]
\centering\small
\caption{Breadth replications: intermediate and terminal broad
adherence and their gap, with scenario-clustered bootstrap CIs;
$\dagger$ marks 95\% intervals that exclude zero.  The single-step
exact sign-flip resolution floor is $p{=}.031$ at six clusters.}
\label{tab:breadth}
\setlength{\tabcolsep}{2.6pt}
\resizebox{\columnwidth}{!}{%
\begin{tabular}{lrrrrrr}
\toprule
& \multicolumn{3}{c}{Single-step (it-helpdesk)} & \multicolumn{3}{c}{English (12 scenarios)} \\
\cmidrule(lr){2-4}\cmidrule(lr){5-7}
Deployment & Int. & Term. & Gap & Int. & Term. & Gap \\
\midrule
ds-v4-flash   & 40.8 & 100.0 & $-59.2^{\dagger}$ & 23.9 & 100.0 & $-76.1^{\dagger}$ \\
ds-v4-pro     & 55.4 & 100.0 & $-44.6^{\dagger}$ & 34.9 & 100.0 & $-65.1^{\dagger}$ \\
glm-5.1       & 60.4 & 100.0 & $-39.6^{\dagger}$ & 45.4 &  97.5 & $-52.1^{\dagger}$ \\
gpt-4.1       & 99.2 & 100.0 & $-0.8$            & 74.7 & 100.0 & $-25.3^{\dagger}$ \\
gpt-4.1-mini  & 61.8 &  97.5 & $-35.7^{\dagger}$ & 40.8 &  62.1 & $-21.3^{\dagger}$ \\
gpt-5.5       & 100.0& 100.0 & $+0.0$            & 81.5 &  92.9 & $-11.4$ \\
qwen3.5-122b-a10b& 46.7 & 100.0 & $-53.3^{\dagger}$ & 38.9 & 100.0 & $-61.1^{\dagger}$ \\
qwen3.5-35b-a3b  & 31.7 & 100.0 & $-68.3^{\dagger}$ & 37.1 &  97.1 & $-60.0^{\dagger}$ \\
qwen3.6-27b   & 54.2 & 100.0 & $-45.8^{\dagger}$ & 30.8 & 100.0 & $-69.2^{\dagger}$ \\
\bottomrule
\end{tabular}
}
\end{table}

The action-cued $D_c{=}0$ positions in both sets show $\approx$0\%
false alarms, consistent with the structural-withholding account of
Appendix~\ref{app:negatives}.

\section{Thinking-mode A/B}
\label{app:thinking}

Could the findings be an artifact of thinking-off configurations?  A
matched two-arm within-batch A/B reran the frozen baseline checkpoints
(CP1--CP5) and the three negative positions of
Appendix~\ref{app:negatives} under paired \emph{thinking-on} vs
\emph{thinking-off} lanes for every deployment whose API exposes a
thinking toggle (10 repetitions; 11{,}520 checkpoints).  Manipulation
checks: thinking lanes must stream nonzero reasoning content in the
admission precheck, and per-row reasoning length is recorded (means
114--218 characters on the verified lanes).  Two endpoints were
excluded at precheck because they inline reasoning into the answer
channel, polluting verdicts; gpt-5.5 does not expose
reasoning at all and its completion sizes are indistinguishable from
thinking-off, so that pair's manipulation is unverifiable and it is
excluded from the conclusions below.

On the four deployments with verified manipulation
(deepseek-v4-pro/-flash, glm-5.1, qwen3.5-35b-a3b), thinking shifts
intermediate broad adherence by $-12.5$ to $+7.5$ points (clustered
bootstrap; largest upper CI $+14.2$) against intermediate--terminal
gaps of $-52$ to $-88$ points: reasoning does not repair
stage-conditioned obligation activation.  Failure mass shifts from
silent omission toward action divergence (one deployment's omission
share drops 36.9\%$\to$20.6\% while divergence rises
21.4\%$\to$39.7\%), mirroring the reminder arm; and on one deployment
OFF-side false alarms worsen by $+12.3$ points (CI $[+6.6,+18.2]$)---%
the precision--recall trade that TriggerBench reports for reasoning
under prospective load \citep{triggerbench2026}, reproduced here on
the control channel.

\section{Matched-obligation bridge: design and scoring}
\label{app:bridge}

The bridge of \S\ref{sec:bridge} runs 12 scenarios $\times$ 5
checkpoints $\times$ 3 repetitions $\times$ 6 arms on nine
deployments (9{,}720 checkpoints).  \textbf{O1} appends a completion
sigil at the final reply only; \textbf{O2} is a stage-triggered
human-readable status sentence; \textbf{O3} is our stage-triggered
machine tag.  Each obligation is run with and without a trailing
reminder, on identical frozen histories.  Scoring uses a separate
deterministic verifier with its own corruption gate.  O2 is reported
under equal-strictness dual views (exact form vs.\ correct-sentence
substring), since 46\% of its strict misses are paraphrases; the
lenient view is the one quoted in the main text (its strict mean is
29.1\%), and the strict view does not change the direction of any
comparison.  The trailing reminder moves the nine-deployment means by
at most $+6.5$ points, with per-deployment shifts from $-9.7$ to
$+20.1$.  Position-matched at the terminal checkpoint the three
obligations reach 99.0\%, 100.0\%, and 97.8\%, and the turns on which
the machine report is absent keep a task-proxy pass rate of 91--100\%.
Table~\ref{tab:bridge} gives the per-deployment values.

\begin{table}[h]
\centering\small
\caption{Matched obligations on nine deployments: adherence at
duty-active checkpoints with the reminder off and on (O1 deferred sigil,
O2 status sentence in the lenient view, O3 machine tag), and the
position-matched O3 values at the terminal checkpoint and at
intermediate checkpoints.  Means are equal-weight over deployments.}
\label{tab:bridge}
\setlength{\tabcolsep}{2.4pt}
\resizebox{\columnwidth}{!}{%
\begin{tabular}{lrrrrr}
\toprule
Deployment & O1 off$\to$on & O2 off$\to$on & O3 off$\to$on & O3@term & O3@int \\
\midrule
deepseek-v4-flash & 100$\to$100 & 38$\to$44 & 56$\to$60 & 100.0 & 41.7 \\
deepseek-v4-pro   & 100$\to$100 & 45$\to$48 & 53$\to$60 & 100.0 & 38.0 \\
glm-5.1           &  97$\to$97  & 31$\to$51 & 56$\to$68 &  94.4 & 42.6 \\
gpt-4.1           & 100$\to$100 & 55$\to$56 & 79$\to$69 & 100.0 & 72.2 \\
gpt-4.1-mini      & 100$\to$100 & 40$\to$56 & 53$\to$69 &  88.9 & 41.7 \\
gpt-5.5           & 100$\to$100 & 86$\to$85 & 88$\to$85 &  97.2 & 85.2 \\
qwen3.5-122b-a10b & 100$\to$100 & 29$\to$39 & 48$\to$53 & 100.0 & 30.6 \\
qwen3.5-35b-a3b   &  94$\to$100 & 28$\to$26 & 36$\to$42 & 100.0 & 14.0 \\
qwen3.6-27b       & 100$\to$100 & 42$\to$48 & 46$\to$43 & 100.0 & 27.6 \\
\midrule
Mean              & 99.0$\to$99.7 & 43.7$\to$50.2 & 57.4$\to$61.0 & 97.8 & 43.7 \\
\bottomrule
\end{tabular}}
\end{table}

\section{Primary and diagnostic metrics and the protocol's task tax}
\label{app:metrics}

\paragraph{Primary metric and diagnostic.}  Because divergence borders
agent-policy adherence (\S\ref{sec:construct}), we report one primary metric
and one diagnostic:
empirical \emph{stage-conditioned control adherence} $\widehat\theta_z$ (clean requires
both the right interaction mode and protocol fulfillment; used in the
main tables) and diagnostic \emph{conditional report validity} $\widehat\phi_z$ (protocol fulfillment
conditional on the absence of an assistant tool call; divergence
excluded from the denominator), using the notation of \S\ref{sec:construct}.
Under the frozen no-tool-call denominator, the descriptive
intermediate--terminal difference in $\widehat\phi_z$ ranges from 1.8 to 83.4
points (Table~\ref{tab:gap}); for example, the duty-blind endpoint is
14.3\% vs.\ 97.7\%, whereas one divergence-dominated endpoint narrows
to 49.0\% vs.\ 54.6\%.  The split separates interaction-mode divergence
from protocol failure after speech is selected, and neither metric alone
tells the full story.

\paragraph{Two metric objections.}  The
primary $\widehat\theta$ gap persists more broadly than the
no-tool-call $\widehat\phi$ gap, confirming that divergence and report validity are
distinct.  Protocol demand has no consistent task-quality effect and is
not required for tool selection, although it amplifies divergence on two
deployments.

\paragraph{Task and protocol outcomes form separate axes.}
We score the archived responses against a frozen scripted-lexical task
proxy whose per-scenario concept tokens were fixed before scoring and
validated at 60/60 agreement in a blinded single-annotator audit archived
in the local evidence package.  Both off-diagonal quadrants are populated
(Table~\ref{tab:axes}); the two rows come from different checkpoint
populations and are not cells of one joint distribution.  The first row is the blind spot of an outcome-only evaluation: the
user-visible turn can remain useful while the runtime channel is absent.
Literal marker-only replies are rarer and occur on two deployments.
These are descriptive diagnostics, not environment-level task outcomes;
they justify treating task behavior and protocol behavior as parallel
rather than substitutable measurements.

\begin{table}[h]
\centering\scriptsize
\caption{Off-diagonal task/protocol outcomes.  Each row has its own
population; the rows are not a joint distribution.}
\label{tab:axes}
\setlength{\tabcolsep}{3pt}
\begin{tabular}{p{0.30\columnwidth}p{0.40\columnwidth}p{0.22\columnwidth}}
\toprule
Quadrant & Population (denominator) & Rate \\
\midrule
Task proxy passed, marker omitted & intermediate checkpoints where the model spoke without the marker & 91--100\% (794/794 on one deployment) \\
Protocol clean, task proxy failed & clarification checkpoint, baseline arm, protocol-clean responses; proxy requires an interrogative form and the required-field token & 4.7--45.4\% on six of seven deployments; none on the seventh \\
\bottomrule
\end{tabular}
\end{table}

\paragraph{Protocol demand has no consistent task tax and amplifies
divergence on two deployments.}  Two objections deserve a direct test: that demanding
the protocol degrades the visible answer, and that the demand is what
pushes models to act instead of speaking---making our headline failure
mode an artifact of our own instrument.  We reran the frozen
checkpoints under two system prompts differing only by the protocol
paragraph (10{,}800 checkpoints, nine deployments).  On task quality,
scored by the frozen proxy over turns where the model chose to speak,
six of nine differences are indistinguishable from zero, two favour the
protocol arm, and one deployment pays a real cost ($-6.5$ points, CI
excluding 0); there is no systematic task tax.  On the
speak-versus-act decision the answer is partly conceding: divergence
exists without any protocol demand at all---two deployments speak on
only 55.3\% and 50.8\% of intermediate checkpoints when no protocol is
required---but requiring it lowers those rates by a further $15.8$ and
$10.3$ points (CIs excluding 0), while six of nine deployments are
unaffected.  Tool-call selection at these positions therefore appears
without any protocol demand; the instrument amplifies it on two of nine,
which bounds how the divergence rates should be interpreted.

\section{Continuation-context interventions: design and per-deployment analysis}
\label{app:affordance}

This appendix gives the design, estimands, and per-deployment results behind the paragraph ``Reports change without task progress'' in \S\ref{sec:factorial-head}.

\paragraph{Setup.}  Figure~\ref{fig:failure-composition} is correlational.  The intervention summarized in \S\ref{sec:affordance} therefore holds the
scripted narrative and mechanically derived gold fixed while changing
visible continuation context.  The first contrast is a
\emph{withdraw-and-delegate treatment bundle}: at the
mid-task post-action checkpoint, gold $(D_c,Z_c^*)=(1,\postobs)$, it both
withdraws the second-step tool and adds a machine-readable field saying
that another system will handle the pending step.  It changes executability and responsibility together, so the
identified contrast is the bundle's total effect on report emission and
on unconditional false completion (estimands in
Appendix~\ref{app:formal}).

\paragraph{Controlled intervention: bundle total effect.}
Pooled over nine deployments (2{,}129 and 2{,}120 successful
checkpoints in the two arms), the bundle increases report emission by
$\widehat\Delta_E^B=75.8\%-30.7\%=+45.1$ points.  On the fixed
successful-response denominator it also increases false completion:
$\widehat\Delta_\mu^B=64.4\%-6.2\%=+58.2$ points: a false
\emph{done} that occurs once in sixteen responses becomes the response
in nearly two of every three, and $\widehat\nu_1=85\%$ of emitted
markers in the bundled arm say \completed{} while the gold is
\postobs.  Per deployment, false completion rises on all nine ($+9.2$
to $+87.6$ points); emission rises on seven ($+17.9$ to $+79.8$) and is
unchanged on two, which switch the value of the reports they already
emit.  Because
the bundle also reassigns responsibility, a model may read \completed{}
as ``nothing remains \emph{for this assistant}'' rather than ``the
user's task is finished''; the effect is completion reporting under
withdrawal, not direct evidence of a false world model.

\paragraph{Fully crossed decomposition.}
To separate the bundle, we ran a $2\times2\times2$ follow-up whose
design and analysis contract were hash-frozen before data collection
(Appendix~\ref{app:repro}); it
that crosses self versus other responsibility, available versus withdrawn
tool declarations, and resolved versus problem feedback cues.  All cells
retain the same unfinished lifecycle truth across 12 scenarios, three
wording families, and eight admitted deployments.  Each reported
factor effect is a main effect averaged over the other two crossed
factors (Appendix~\ref{app:formal}).

\begin{figure}[h]
\centering
\includegraphics[width=\columnwidth,keepaspectratio]{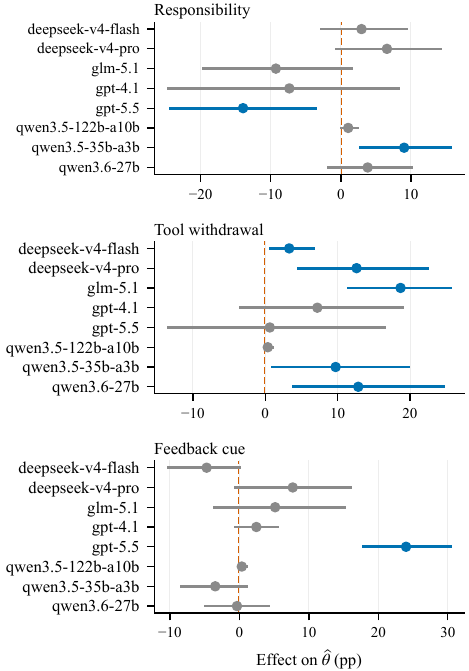}
\caption{Per-deployment effects in the controlled $2\times2\times2$
decomposition: responsibility (other $-$ self), tool withdrawal
(withdrawn $-$ available), and feedback cue (problem $-$ resolved).
Tool withdrawal is non-negative on all eight deployments;
responsibility and cue effects are heterogeneous.}
\label{fig:factorial-forest}
\end{figure}

Tool withdrawal is the only factor whose end-to-end effect is
non-negative on all eight deployments ($+0.3$ to $+18.7$ points; two
effects lie within one point of zero); five of the eight
scenario-cluster intervals exclude zero and six of eight keep the same
sign in all three wording families.  Its two fixed-denominator
diagnostics are positive in all deployment$\times$wording checks.
Responsibility also raises both diagnostics, but its end-to-end effect
is mixed (two of eight intervals exclude zero).  Both directions run
through emission: withdrawal makes the model report more often, and the
added reports split between the correct \postobs{}, which raises
$\widehat\theta$, and \completed{}, which raises false completion.  Cue and
interaction effects are heterogeneous
(Figure~\ref{fig:factorial-forest}; full ranges in Appendix
Table~\ref{tab:factorial}).  Figure~\ref{fig:factorial-forest} shows
percentage-point effects on $\widehat\theta$ with scenario-cluster
bootstrap 95\% CIs over 12 scenarios; blue intervals strictly exclude
zero, while a bound exactly at zero remains gray.  Deployment order is
alphabetical and is not a ranking; Appendix Table~\ref{tab:factorial}
adds full ranges, emission, and false-completion columns.  Hence the
follow-up identifies behavioral effects of explicit fields and tool
declarations, not a universal effect or an internal model state.
Conditional report validity remains secondary because treatment changes
which turns enter its selected denominator.

One bounded reading is consistent with all of these results: emitting
a runtime-facing report competes with continuing to act.  Embodied
systems show the same pattern, continuing to act after a hidden goal is
satisfied rather than issuing the required terminal report
\citep{vigil2026}; and a direct state question often recovers an
omitted report but does not repair an incorrect one
(Appendix~\ref{app:repair}).

\section{Follow-up Repair Diagnostics}
\label{app:repair}

These interventions concern distinct sampled populations and are not
randomized arms of one experiment.  On 1{,}677 responses from three
deployments that contained useful natural language but omitted the
lifecycle tag, replay plus a short continuation recovered the correct
tag on 756/1{,}677 responses (45.1\%), compared with 44.2\% without the
newline.  When a tag appeared, its value was almost always correct
(756/763), but 921 responses still produced no correct tag.

An isolated state-classification request produced a state signal on
100\% of 2{,}654 requests and selected the correct value on 98.12\%
(all gold states were \needsinfo{}).  On the 1{,}677 source-paired rows,
the isolated question recovered 896 rows that continuation missed,
while continuation recovered no row the isolated question missed.

The same question did not repair already incorrect state reports.  In a
separate \postobs{} population, six deployments passed the
pre-registered 95\% lane-completion gate; across 820 successful
verbatim follow-ups, all 820 again reported \completed{}.  Removing the
previous erroneous marker produced only 0--0.6\% correct responses per
deployment.  Two additional deployments missed the completion gate and
remain infrastructure-missing observations; one deployment produced no data
after repeated precheck failures.

\begin{table}[h]
\centering\small
\caption{Repair diagnostics.  The populations differ, so the rows are
not a randomized between-group effect.}
\label{tab:repair-boundary}
\setlength{\tabcolsep}{3pt}
\begin{tabular}{llrr}
\toprule
Population & Intervention & $n$ & Correct \\
\midrule
Tag omitted & Replay + newline & 1{,}677 & 45.1\% \\
Tag omitted & Isolated question & 2{,}654 & 98.12\% \\
Wrong state & Same question & 820 & 0.0\% \\
\bottomrule
\end{tabular}
\end{table}

The result separates two diagnostics: a direct question can often
recover an omitted report in the sampled omission population, but it is
not an independent authority for lifecycle state and does not repair an
already incorrect judgment in the sampled wrong-state population.

\section{Extended related work}
\label{app:related}

\paragraph{Coverage of the survey.}  The comparison in this appendix
draws on a structured scan of the agent-evaluation literature published
between August 2025 and August 2026, run along four axes---termination
and self-reporting; multi-turn instruction following and prospective
memory; trajectory and process evaluation; and structured control
signals and runtime systems---yielding 62 candidate papers.  Venue
claims were verified at the publisher or preprint source rather than
taken from secondary listings; where no venue could be confirmed, the
work is cited as a preprint.  This is a recency sweep and does not
replace the older foundational work cited in \S\ref{sec:related}.  A
second scan along a progress-self-estimation axis added three works,
one of which predates the window
\citep{bishop2024latent,repro2026,park2026mirage}.

\begin{table*}[t]
\centering\scriptsize
\caption{Construct map.  The rows can share surface syntax while differing
in what is evaluated, when the obligation applies, where truth comes from,
and who consumes the output.  StageIF targets the final row rather than
treating lifecycle reporting as generic instruction following.}
\label{tab:construct-map}
\setlength{\tabcolsep}{3pt}
\begin{tabular}{p{0.17\textwidth}p{0.19\textwidth}p{0.20\textwidth}p{0.20\textwidth}p{0.17\textwidth}}
\toprule
Construct & Primary object & Activation condition & Truth / reference source & Primary consumer \\
\midrule
Instruction following & response constraint & instruction-dependent & gold answer or checker & user / evaluator \\
Structured output & serialized response & usually each evaluated response & schema plus content reference & application / evaluator \\
Prospective memory & deferred action & future cue or deadline & requested future duty & user / environment \\
Process or state evaluation & action or trajectory state & milestone or terminal event & environment or trace & evaluator \\
Latent-state estimation & model-estimated progress and completion & every step & human annotation & agent's own planner \\
Lifecycle reporting (StageIF) & model-reported control state & runtime-derived checkpoint & trusted runtime state & runtime control logic \\
\bottomrule
\end{tabular}
\end{table*}

\paragraph{Multi-turn instruction following.}  Multi-IF shows
instruction adherence degrades over turns \citep{he2024multiif};
MultiChallenge disentangles instruction retention, context allocation,
and reasoning in realistic multi-turn settings
\citep{sirdeshmukh2025multichallenge}; StructFlowBench treats
cross-turn structural dependencies as first-class
\citep{li2025structflow}; EvolIF tracks dynamically evolving
constraints \citep{wang2026evolif}.  All evaluate human-readable
answers under general constraints.

\paragraph{Structured output.}  SchemaBench and StructEval show
syntactic validity is itself nontrivial
\citep{geng2025jsonschemabench,gu2025structeval}; the Format Tax shows
the demand for format, more than decoder constraints, shifts behavior
\citep{lee2026formattax}.  These are single-turn with a fixed,
always-applicable schema, so they cannot express an obligation that is
active at some turns and forbidden at others.

\paragraph{Progress self-estimation.}  Latent-state estimation prompts
a UI agent to estimate performed actions, progression, mistakes, and
completion at every step and scores the estimates against human
annotation on a 40-task subset, reporting 87.4\% for progression and
97.3\% for completion; the estimates feed the agent's own planner
\citep{bishop2024latent}.  The reporting duty there is always on, truth
is annotated rather than derived from environment state, and
correctness is binary, so an omitted report and a wrong stage value are
not distinguished and accuracy is not compared across stages.  Under a
stage-derived duty with environment-derived truth, the two older
deployments in \S\ref{sec:tau2} report the mid-task stage correctly on
5.8--11.5\% of checkpoints, which shows how much the measurement design
moves the answer.  RePro trains agents to generate progress percentages,
states that per-step progress lacks ground truth in outcome-based
tasks, and finds in a pilot that online progress prompting hurts task
performance \citep{repro2026}.  A preregistered pilot on one
long-running agent loop finds that the agent claimed improvement in
every cycle while most cycles showed no measured gain
\citep{park2026mirage}.

\paragraph{Stateful tool agents and process evaluation.}  ToolSandbox
evaluates stateful execution with intermediate milestones
\citep{lu2025toolsandbox}; BFCL and ACEBench extend function-calling
evaluation to multi-turn agentic settings
\citep{patil2025bfcl,chen2025acebench}; DialogTool decomposes the
tool-use lifecycle \citep{liu2025dialogtool}.  Step- and
trajectory-level methods score intermediate actions or whole
trajectories rather than only final outcomes
\citep{wang2025steca,zhang2026planreward,proxystate2026};
SOPBench evaluates whether agents follow standard operating procedures
when acting \citep{sopbench2025}, and OctoBench whether persistent
scaffold rules survive long interactions \citep{octobench2026}.  All
target actions, policies, or task milestones, leaving the
model-to-orchestrator communication channel unmeasured.  AgentIF
evaluates conditional and tool constraints inside realistic agentic
prompts \citep{agentif2025}, but not obligations activated by
interaction stage.  Inter-agent protocol benchmarks compare
communication protocols \emph{between} agents
\citep{protocolbench2025}---a different channel from the
model-to-runtime lifecycle signal studied here.

\paragraph{Position within prospective memory.}  Beyond the reminder
contrast reported in \S\ref{sec:factorial}, our work extends this line
from single-response formatting duties to a multi-turn,
stage-activated, machine-consumed protocol with $D_c{=}0$ positions,
and decomposes failure into applicability, state selection, and
realization.  PM-Bench's Virtual Week schedule carries event- and
time-based tasks with latent-channel monitoring; TriggerBench measures
proactive recall and false alarms.  Both target semantic-level
proactive behavior; neither measures stage-value selection or
realization on a machine-consumed channel.

\section{Runtime-owned gate replay}
\label{app:gateab}

The 336 archived live trajectories are replayed under two continuation
gates: \emph{model-emitted}, where a missing marker kills the session
and a \texttt{COMPLETED} marker stops it, and \emph{runtime-owned},
where the lifecycle is derived from environment observables.  Because
gate decisions depend only on the trajectory prefix, truncating at the
first stop event reproduces the gated outcome distribution exactly with
zero provider calls.  The model-emitted gate yields 114/336 correct
terminations, 62 dead ends, and 8 premature stops; the runtime-owned
gate yields 164/336 with both failure classes eliminated by
construction.  The residual incompletes are behavioral and shared by
both gates, and zero sessions finish the environment without
signaling.  Replay rules are in the artifact repository
(Appendix~\ref{app:repro}).

\section{Runtime gate comparison: design and results}
\label{app:gate}

This appendix gives the validating rollout introduced in
\S\ref{sec:realenv}.  Its marker gate ends the session when a report is
missing, so outcomes concern task completion, not just report accuracy.
The comparison is separate from the interventions in
\S\ref{sec:factorial-head}.  The no-user setting in Appendix~\ref{app:tau2-deciles}
also consumes a model-generated stop signal, with 63 of 114 episodes
ending at zero reward; that count alone establishes neither false
completion claims nor the cost of the stop rule.  The rollout below
compares reporting-and-termination configurations on scripted tasks.

\paragraph{Setup.}  Scripted checkpoints isolate behavior but not system cost, so we use a
validating rollout that rejects structurally invalid actions and
enforces cross-step referential integrity, with task success derived
from environment state.  Across 12 scenarios, six repetitions, and eight
deployments, we compare the marker gate B1, the no-tool-call gate B2, and
runtime-owned control A.  B1 and B2 match user-message templates, but
the implementation adds reporting instructions only for the marker
gate.  Matching user messages therefore does not isolate the gate
rule from the reporting protocol.  A additionally changes the
continuation pathway (implementation and rejection checks in
Appendix~\ref{app:realenv}).  Each trajectory ends
in one of four outcomes: \emph{task ok} (both steps executed validly,
read from state); \emph{dead end} (a natural-language turn carried no
marker and the gate ended the session); \emph{premature stop} (the gate
stopped on a \completed{} marker while the task was unfinished); or
\emph{stall} (the session returned to a user who had nothing to add).

Table~\ref{tab:realenv} covers 2{,}304 trajectories over eight
deployments and omits turn-cap and transport terminations (0.2--1.6\%;
Appendix~\ref{app:realenv}).  Dead ends and premature stops are
structurally impossible without a marker to read; the two runtime-owned
rows differ only in whether the re-prompt carries the default-value
authorisation that the other gates' user turn does.  B2 removes every
dead end but stalls 45.5\% of trajectories.

\begin{table}[h]
\centering\small
\caption{Termination design and outcome in a validating environment.
Panel~A matches user messages but not all model inputs; it compares
reporting-and-termination configurations, not isolated gate effects.
Panel~B additionally varies the controller, continuation speaker, or
authorisation.}
\label{tab:realenv}
\setlength{\tabcolsep}{3pt}
\resizebox{\columnwidth}{!}{%
\begin{tabular}{lrrrr}
\toprule
\multicolumn{5}{l}{\emph{Panel A: model-signaled configurations}} \\
Gate & Task ok & Dead end & Prem. & Stall \\
\midrule
Marker (B1)      & 40.8\% & \textbf{22.7\%} & 1.6\% & 33.3\% \\
No-tool-call (B2)& 53.8\% & 0\% & 0\% & 45.5\% \\
\midrule
\multicolumn{5}{l}{\emph{Panel B: runtime-owned configurations}} \\
\midrule
A, unmatched & 40.8\% & 0\% & 0\% & 59.0\% \\
A, matched & \textbf{81.1\%} & 0\% & 0\% & 18.6\% \\
\bottomrule
\end{tabular}
}
\end{table}

\paragraph{A 13-point difference between tested configurations.}
Task success rises from 40.8\% under the marker gate (B1) to 53.8\%
under the no-tool-call gate (B2).  The observed configuration contrast
(defined in Appendix~\ref{app:formal}) is $+13.0$
points (scenario-clustered CI $[+7.1,+19.4]$;
Table~\ref{tab:realenv}).  This contrast includes the marker gate's
strict response to an absent report: it ends the session rather than
waiting for a repair; it does not separate that rule from the reporting
instructions.  The task-success difference is heterogeneous
(Appendix~\ref{app:realenv}, Table~\ref{tab:realenv-dep}): the B2$-$B1 task-success difference is
positive on seven deployments ($+4$ to $+68$ points) and negative on
one ($-25$), and 110 of the 131 dead ends fall on two deployments.
Omission-heavy deployments incur dead ends, whereas divergence-heavy
deployments often produce no natural-language turn for the marker gate
to reject; this deployment-level association is exploratory, not
predictive.

\paragraph{Missing reports stop tasks before they finish.}
The marker gate ends 22.7\% of trajectories because a marker is
missing, whereas premature completion claims terminate only 1.6\%.
Missing-marker stops are more frequent in this arm, but these outcome
frequencies do not identify how much of the between-configuration
success difference each failure causes.  No gate rejects a
trajectory after both steps complete.  The marker gate instead stops
sessions before completion, typically at clarification: the loss is
unfinished work, not discarded completed work.

\paragraph{Surviving terminal reports are a selected sample.}
Duty-active intermediate adherence under the marker gate remains
4.2--38.0\% across the eight deployments, so the deficit persists in
an environment that validates actions.  All 235 surviving marker-gate
terminal checkpoints (one per task-ok trajectory, 235/576) were CLEAN.
But those trajectories first had to survive the earlier marker-gated
turns.  The terminal rate is not guaranteed by definition, nor is it
comparable to the scripted-checkpoint terminal estimate; it cannot
establish a rollout stage gap (Appendix~\ref{app:realenv}).

\paragraph{Configuration comparison.}  The runtime-owned arm reaches 81.1\% with the matched re-prompt, 40.3
points above the unmatched wording, but also changes the continuation
pathway, so it is a configuration comparison, not evidence
that one architecture is universally superior; the bounded conclusion
is that a model-generated lifecycle report is useful as diagnostic
evidence but unsafe as the sole authority when the corresponding state
can be derived from trusted runtime events.  An independent audit
points the same way: false-success rates fall by an order of magnitude
in the one domain where an independent simulator could verify state
\citep{advani2026}.

\section{Validating-environment construction and its audit}
\label{app:realenv}

\paragraph{Outcome coverage in Table~\ref{tab:realenv}.}  The four
outcome columns are not exhaustive: each arm also contains turn-cap and
transport terminations, excluded from the table because neither is a
property of the termination rule under test.  Per arm ($n{=}576$ each):
marker 6 turn-cap $+$ 3 transport; no-tool-call 1 $+$ 3; runtime-owned
unmatched 0 $+$ 1; runtime-owned matched 2 $+$ 0.  Rows therefore sum
to 98.4--99.8\%.  Table~\ref{tab:realenv-dep} gives the per-deployment
outcomes of the message-matched pair.

\begin{table}[h]
\centering\small
\caption{Per-deployment outcomes of the message-matched gates
(counts out of 72 trajectories each; turn-cap and transport
terminations omitted).  B1 is the marker gate, B2 the no-tool-call
gate.  Deployments are alphabetical, not ranked.}
\label{tab:realenv-dep}
\setlength{\tabcolsep}{2.6pt}
\resizebox{\columnwidth}{!}{%
\begin{tabular}{lrrrrrrr}
\toprule
 & \multicolumn{4}{c}{B1 marker} & \multicolumn{2}{c}{B2 no-tool-call} & \\
\cmidrule(lr){2-5}\cmidrule(lr){6-7}
Deployment & ok & dead & prem. & stall & ok & stall & B2$-$B1 (pts) \\
\midrule
deepseek-v4-flash & 18 & 41 & 0 & 13 & 26 & 46 & $+11$ \\
deepseek-v4-pro   & 25 &  5 & 0 & 42 & 31 & 40 & $+8$ \\
glm-5.1           & 52 &  0 & 5 & 15 & 55 & 17 & $+4$ \\
gpt-4.1           & 22 &  0 & 1 & 48 & 30 & 41 & $+11$ \\
gpt-4.1-mini      & 17 &  5 & 1 & 47 & 25 & 46 & $+11$ \\
qwen3.5-122b-a10b & 50 &  0 & 0 & 17 & 32 & 40 & $-25$ \\
qwen3.5-35b-a3b   &  2 & 69 & 0 &  1 & 51 & 20 & $+68$ \\
qwen3.6-27b       & 49 & 11 & 2 &  9 & 60 & 12 & $+15$ \\
\bottomrule
\end{tabular}}
\end{table}

\paragraph{What the environment checks.}  Each scenario is backed by a
typed state store.  Tool names are dispatched, not ignored; step-one
requires every declared field, non-empty, with the field the user had
to supply matched against a per-scenario acceptance pattern; step-two
requires the identifier minted by step-one, verbatim.  Task success is
the conjunction of both steps having executed validly, read from state.
Identifiers a live agent cannot know---an opaque order or account
number never spoken by the user nor returned by a tool---are resolved
from session context rather than demanded, since requiring them would
measure clairvoyance.

\paragraph{Construction record.}  The environment was rebuilt three
times after defects found in scheduling back-off, in the time-field
check, and in its replacement; the runtime-owned gate was run twice
with one wording change (the 40.3-point difference reported in
Appendix~\ref{app:gate}); and a second session audited the environment
twice for false kills.  Each cycle, the rerun rule, and both audit
corrections are recorded in the artifact repository
(Appendix~\ref{app:repro}).

\section{Reproducibility}
\label{app:repro}
Stimuli (360, hash-frozen), scenario scripts, state machine, verifiers
(both strictness variants), corruption-gate tests, per-request ledgers,
and raw per-checkpoint outcomes are archived with SHA-256 manifests, and
provider-free scripts recheck archive hashes, cardinalities, the
paper's primary metric cells, and the scenario-level exact tests from
stored responses.  This establishes pipeline and metric
recomputability, not behavioral replication on a public model.  The
implementation record that this paper does not reproduce---exact prompt
generators and frozen stimuli per study family, a worked
case-to-score trace, the validating environment's defect-and-rerun
cycles and independent audits, verifier forensics, gate-replay rules,
the framework survey's collection protocol, and all figure
generators---is kept in a companion artifact repository, private
during review and released subject to privacy, licensing, owner, and
venue review.

\end{document}